\documentclass[english,superscriptaddress,amsmath,twocolumn,amssymb,aps,longbibliography]{revtex4-1}
\usepackage{babel}
\usepackage{xcolor}
\usepackage{array}
\usepackage{booktabs}
\usepackage{bm}
\usepackage{multirow}
\usepackage{graphicx}
\usepackage{hyperref}
\hypersetup{colorlinks,citecolor=blue,linkcolor=blue}

\begin{document}
	\title{Topological bosonic Bogoliubov excitations with sublattice symmetry}
	\author{Ling-Xia Guo}
	\affiliation{School of Physics and Institute for Quantum Science and Engineering, Huazhong University of Science and Technology, Wuhan 430074, China}
	\author{Liang-Liang Wan}
	\email{wanlianghust@163.com}
	\affiliation{School of Physics and Institute for Quantum Science and Engineering, Huazhong University of Science and Technology, Wuhan 430074, China}
	\author{Liu-Gang Si}
	\affiliation{School of Physics and Institute for Quantum Science and Engineering, Huazhong University of Science and Technology, Wuhan 430074, China}
	\author{Xin-You L\"{u}}
	\affiliation{School of Physics and Institute for Quantum Science and Engineering, Huazhong University of Science and Technology, Wuhan 430074, China}
	\author{Ying Wu}
	\affiliation{School of Physics and Institute for Quantum Science and Engineering, Huazhong University of Science and Technology, Wuhan 430074, China}
\begin{abstract}
Here we investigate the internal sublattice symmetry, and thus the enriched topological classification of bosonic Bogoliubov excitations of thermodynamically stable free-boson systems with non-vanishing particle-number-nonconserving terms. Specifically, we show that such systems well described by the bosonic Bogoliubov-de Gennes Hamiltonian can be in general reduced to particle-number-conserving (single-particle) ones. Building upon this observation, the sublattice symmetry is uncovered with respect to an excitation energy, which is usually hidden in the bosonic Bogoliubov-de Gennes Hamiltonian. Thus, we obtain an additional topological class, i.e., class AIII, which enriches the framework for the topological threefold way of free-boson systems. Moreover, a construction is proposed to show a category of systems respecting such a symmetry. For illustration, we resort to a one-dimensional (1D) prototypical model to demonstrate the topological excitation characterized by a winding number or symplectic polarization. By introducing the correlation function, we present an approach to measure the topological invariant. In addition, the edge excitation together with its robustness to symmetry-preserving disorders is also discussed.
\end{abstract}
\maketitle

\section{Introduction}
Topological states of matter beyond the Landau paradigm attract growing attentions due to their fundamental novelty and great prospect of applications~\citep{Hasan2010RevModPhys, Qi2011RevModPhys, Chiu2016RevModPhys, Wen2017RMP}. In particular, a mature framework has been established for classifying ground states of topological insulators and superconductors of free fermions based on the (internal) time-reversal, particle-hole and chiral symmetries, namely, Altland-Zirnbauer (AZ) classification~\citep{Schnyder2008PRB, Kitaev2009AIP, Ryu2010NJP, Chiu2016RevModPhys}. In contrast to fermionic systems where all states are equally occupied within bands below the Fermi level due to the fermionic nature, the non-interacting bosons condense to the lowest-energy mode (i.e., ground state) and, hence, does not exhibit any topologically nontrivial phases. However, people have found many exotic topological phenomena of wave functions of bosonic excitations in photonic materials~\citep{Haldane2008PhysRevLett, Wang2008PhysRevLett, Wang2009Nature, Fu2011APL, Hafezi2013NatPhotonics, Wang2013PhysRevB, Rechtsman2013Nature, Skirlo2014PhysRevLett, Lu2014NatPhoto, Skirlo2015PhysRevLett, Wu2015PhysRevLett, Siroki2017PhysRevB, Mittal2019PRL, Ozawa2019RevModPhys, Smirnova2020APR, Price2022JPPhoto, Lan2022RP, Tang2022LPR}, magnonic systems~\citep{Matsumoto2011PRL, Zhang2013PRB, Shindou2013PRB, Chisnell2015PRL, RoldanMolina2016NJP, Nakata2017PRB, McClarty2018PRB, Yao2018NatPhys, Kondo2019PysRevB, Diaz2019PRL, Zhu2021SciAdv, Mook2021PRX, McClarty2022ARCMP, Massarelli2022PRB}, optical lattices loaded with cold bosonic atoms~\citep{Jaksch1998PhysRevLett, Kawaguchi2012PR, Aidelsburger2014NatPhys, Furukawa2015NJP, Xu2016PhysRevLett, Di_Liberto2016PRL, GonzalezCuadra2020PRL, Wang2020PRA, Cardarelli2023QST}, etc. For instance, the spontaneous symmetry breaking can result in the emergence of topological superfluid in cold-atom optical lattices~\citep{Xu2016PhysRevLett, Luo2018PRA, Huang2021PRA}. The squeezed light can induce the inelastic edge-state transport and even topological quantum fluctuation in photonic systems~\citep{Peano2016NatCommun, Peano2016PRX}.

Among them, the intriguing topological effects of excitations are determined by the bosonic Bogoliubov-de Gennes (BdG) Hamiltonian, which describes the system emergent from (weakly) interacting bosons under the mean-field approximation~\citep{Ring1980, Blaizot1986}. Such a system, dubbed ``free boson'', manifests the pseudo-unitary structure inherently~\citep{Gurarie2002PRL, Gurarie2003PRB, Shindou2013PRB}, which has no counterpart in the AZ classification. To understand it, a framework of characterizing topological (high-energy) excitations of free-boson systems in the thermodynamic-stability regime, which is distinct from the classification of phases (i.e., ground-state physics), has been built, known as the topological threefold way~\citep{Lu2018arXiv, Kawabata2019PRX, Lein2019PRB, Zhou2020JPA, Xu2020PRB, Chaudhary2021PRB}. Remarkably, this new framework shows the topological triviality of one-dimensional (1D) bosonic Bogoliubov excitations. It is a great pity for both theoretical and experimental researchers that the simplest subject has to be crossed off the shortlist. Such a situation prompts us to doubt: Is the physics of the topological bosonic Bogoliubov excitation for the thermodynamically stable free-boson system in one dimension ruled out completely? Or, is there any other internal symmetry enabling the topological bosonic excitation? Answering these questions will inspire us to deepen our understanding of bosonic topological physics and to explore novel topological phenomena.

In this work, we study the topological bosonic Bogoliubov excitation in the free-boson system with particle-number-nonconserving terms under the thermodynamic-stability regime. Specifically, we show that the particle-number-nonconserving bosonic BdG Hamiltonian can be reduced to a particle-number-conserving (single-particle) one via a pseudo-unitary transformation if its Hamiltonian is positive definite. This reduction enables us to map the free-boson system to the one in the AZ classification without the particle-hole and chiral symmetries. In terms of the absence or presence of time-reversal symmetry, such systems are classified by the topological threefold way, i.e., class A, AI and AII. It is worth noting that the mapping adopted here preserves those symmetries concerned at the single-particle level in the topological classification. We prove that this mapping adopted from Ref.\,\citep{Zhou2020JPA} even holds in the dynamical-stability regime (that is, the modes of the Hamiltonian exhibit strictly bounded motion in time), and it is topologically equivalent to that of Ref.~\citep{Chaudhary2021PRB} although their formulas are different from each other. More importantly, based on the single-particle Hamiltonian, we point out the existence of sublattice symmetry, which is usually hidden in the bosonic BdG Hamiltonian and mathematically similar to chiral symmetry at the single-particle level for free-fermion systems. Such an internal symmetry supplies an additional topological class AIII and it implies the existence of topological bosonic Bogoliubov excitation in one dimension. Through a construction, we also show a category of free-boson systems which physically have the sublattice symmetry.

To illustrate the topological bosonic Bogoliubov excitation, we provide a 1D prototypical bosonic model. Utilizing the pseudo-unitary transformation, we show that such a model has sublattice symmetry, and the associated band topology is characterized by the winding number or, equivalently, the symplectic polarization. To measure the nontrivial topology, an approach is proposed by distinguishing the monotonicity of envelope of the resonance reflected by the correlation function in frequency domain. Explicitly, the resonance envelope behaves monotonically as increasing the frequency for the nontrivial topological number, while it does not for the trivial number. Moreover, we present the edge excitation that emerges when the open boundary condition is applied. It can be diagnosed via the midgap peak of the resonance given by the correlation function. As a topological effect, the edge excitation is robust against the symmetry-preserving disorders.

The paper is organized as follows. In Sec.\,\ref{sec:Sym}, we introduce the bosonic BdG Hamiltonian for the free-boson system and demonstrate that it can be reduced to a particle-number-conserving (single-particle) Hamiltonian by a pseudo-unitary transformation. In Sec.\,\ref{sec:Hidden_Sym_Top}, we prove the mapping adopted here preserves the symmetries concerned in the context of AZ classification and, then, obtain the standard threefold way for the bosonic system. Importantly, we unveil the hidden sublattice symmetry and obtain the enriched topological classification, i.e., class A, AI, AII and AIII. A construction for the bosonic BdG Hamiltonian with the sublattice symmetry is also discussed. After that, we make comparisons with the existing literature. Section~\ref{sec:ExI} resorts an example to illustrate the topological bosonic Bogoliubov excitation in one dimension. The topological-invariant measurement together with the edge excitation is subsequently discussed in detail. We also show the robustness to symmetry-preserving disorders. Finally, the conclusion is given in Sec.~\ref{sec:Conclusion}.

\section{Deconstruction of bosonic Bogoliubov-de Gennes Hamiltonian \label{sec:Sym}}
Generally, a free-boson system represents the mean-field description of the (weakly) interacting bosonic system, which ubiquitously exists in photonic systems~\citep{Ozawa2019RevModPhys, Smirnova2020APR}, cold bosonic atoms~\citep{Pethick2008, Zhai2021}, magnonic crystals~\citep{Kondo2020PTEP}, exciton-polaritons~\citep{Bardyn2016, Karzig2015PhysRevX, Bleu2016PRB}, etc. And it is well described by the bosonic BdG Hamiltonian in a $d$-dimensional ($d$D) spatial space
\begin{equation}
\hat{H}= \sum_{i,j=1}^{N}\left[\hat{a}_{i}^{\dagger}K_{ij}\hat{a}_{j}\\
+\frac{1}{2}\left(\hat{a}_{i}^{\dagger}M_{ij}\hat{a}_{j}^{\dagger}+{\rm H.c.}\right)\right],
\label{eq:H1}
\end{equation}
where $\hat{a}_{i}$ and $\hat{a}_{i}^{\dagger}$ ($i=1,2,\ldots,N$) are the bosonic annihilation and creation operators. Here $N$ denotes the possible physical degrees of freedom in bosonic systems, such as orbitals, (pseudo-)spins, sublattices, etc. The $N$-dimensional matrix $K$ ($K=K^{\dagger}$) describes the single-particle part in the free-boson system and usually determines the energy dispersion. The particle-number-nonconserving term $\frac{1}{2}\sum_{i,j=1}^{N}\hat{a}_i^\dagger M_{ij}\hat{a}_j^\dagger+{\rm H.c.}$ ($M=M^{T}$) arises from the interaction under the mean-field approximation. By adopting the Nambu spinor ${\hat{\Phi}} = ({\hat{a}_{1}},\ldots,{\hat{a}_{N}}, {\hat{a}_{1}^{\dagger}},\ldots, {\hat{a}_{N}^{\dagger}})^{T}$, we obtain $\hat{H} = \frac{1}{2}  \hat{\Phi}^{\dagger} H\hat{\Phi} - \frac{1}{2}{\rm Tr}K$, where
\begin{equation}
H=\left(\begin{array}{cc}
K & M\\
M^{*} & K^{T}
\end{array}\right)
\label{eq:H2}
\end{equation}
is the BdG Hamiltonian in the matrix form.

Suppose that the free-boson system has the translational invariance and satisfies the periodic boundary condition. The bosonic BdG Hamiltonian (\ref{eq:H1}) can be transformed into the momentum space, and rewritten as $\hat{H}=\frac{1}{2}\sum_{\bf k}\hat{\Phi}^\dagger_{\bf k}H({\bf k})\hat{\Phi}_{\bf k}-\frac{1}{2}{\rm Tr}K$. Here $\hat{\Phi}_{\bf k} = (\hat{a}_{{\bf k}1}, \ldots, \hat{a}_{{\bf k}\tilde{N}}, \hat{a}^\dagger_{-{\bf k}1}, \ldots, \hat{a}^\dagger_{-{\bf k}\tilde{N}})^T$, and the Hamiltonian matrix in the momentum space is given by
\begin{equation}
	H({\bf k})=\left(\begin{array}{cc}
		K({\bf k}) & M({\bf k})\\
		M^{*}(-{\bf k}) & K^{T}(-{\bf k})
	\end{array}\right).
	\label{eq:H_momentum}
\end{equation}
Here $M(-{\bf k})=M^T({\bf k})$ and ${\bf k}=(k_1,k_2,\ldots,k_d)$ is the crystal momentum in the $d$D space. In this case, the dimension of the BdG Hamiltonian matrix (\ref{eq:H2}) $2N$ is reduced to $2\tilde{N}$ by eliminating the spatial degrees of freedom. The spinor $\hat{\Phi}_{\bf k}$ satisfies the following bosonic commutation relations
\begin{align}
	&[\hat{\Phi}_{{\bf k}i},\hat{\Phi}_{{\bf k}'j}]={\rm i}\delta_{{\bf k}{\bf k}'}(\tau_2)_{ij}, \ \ \ \tau_2=\left(\begin{array}{cc}
		& -{\rm i} I_{\tilde{N}}\\{\rm i}I_{\tilde{N}} & \\
	\end{array}\right), \label{eq:CCR1}	\\
	&[\hat{\Phi}_{{\bf k}i},\hat{\Phi}_{{\bf k}'j}^{\dagger}]=\delta_{{\bf k}{\bf k}'}(\tau_{3})_{ij},  \ \ \ \tau_{3}=\left(\begin{array}{cc}
		I_{\tilde{N}} & \\ & -I_{\tilde{N}}
	\end{array}\right), 
	\label{eq:CCR2}
\end{align}
with $I_{\tilde{N}}$ being a $\tilde{N}$-dimensional identity matrix. Hereafter, $\tau_{1}={\rm i}\tau_{3}\tau_{2}$. $\tau_3$ is the indefinite metric of the underlying Krein space $\mathbb{J}$~\citep{Rossignoli2005PhysRevA, Gohberg2005, Lein2019PRB}, which is the tensor product of the non-spatial physical space $\mathbb{C}^{\tilde{N}}$ with the particle-hole degree of freedom $\mathbb{C}^{2}$~\citep{Altland1997PhysRevB, Peano2018JMP}.

The quadratic Hamiltonian on the Krein space can be transformed to the one acting on the symplectic space. To see this, instead of writing the quadratic Hamiltonian $\hat{H}$ using the Nambu spinor, one can use the Hermitian canonical coordinates and momenta, $\hat{x}_{{\bf k}i}=(\hat{a}_{{\bf k}i}+\hat{a}_{-{\bf k}i}^\dagger)/\sqrt{2}$ and $\hat{p}_{{\bf k}i}=(\hat{a}_{{\bf k}i}-\hat{a}_{-{\bf k}i}^\dagger)/\sqrt{2}{\rm i}$ [$(\hat{a}_{{\bf k}i})^\dagger=\hat{a}^\dagger_{-{\bf k}i}$], which are also called quadratures. In the quadrature representation, $\hat{\xi}_{\bf k}=(\hat{x}_{{\bf k}1},\ldots,\hat{x}_{{\bf k}\tilde{N}},\hat{p}_{{\bf k}1},\ldots,\hat{p}_{{\bf k}\tilde{N}})^T$, the Hamiltonian $\hat{H}$ can be rewritten as $\hat{H}=\frac{1}{2}\sum_{\bf k}\hat{\xi}_{\bf k}^T R({\bf k})\hat{\xi}_{\bf k}-\frac{1}{2}{\rm Tr}K$, and it describes a system of $\tilde{N}$ mutually coupled quantum bosonic oscillators. The Hermitian operator $\hat{\xi}_{\bf k}$ satisfy the commutation relation $[\hat{\xi}_{{\bf k}i},\hat{\xi}_{{\bf k}'j}]=\delta_{{\bf k}{\bf k}'}(\tau_{2})_{ij}$. The quadrature and bosonic representation are related by $\hat{\xi}_{\bf k}=G\hat{\Phi}_{\bf k}$ and $R({\bf k})=GH({\bf k})G^\dagger$, where
\begin{equation}
G=\frac{1}{\sqrt{2}}\left(\begin{array}{cc}
I_{\tilde{N}} & I_{\tilde{N}}\\
-{\rm i} I_{\tilde{N}} & {\rm i}I_{\tilde{N}}
\end{array}\right)
\label{eq:CCR_q}
\end{equation}
is a unitary matrix. Thus, the Krein space $\mathbb{J}$ is then transformed into the space with the symplectic structure ${\rm i}\tau_{2}$, i.e., symplectic space, and they are related by the unitary transformation $G$.

Here we are interested in the thermodynamic-stability regime where the system has a ground state and the energy has a lower bound, $E\geq0$~\citep{Dderezinski2017JMP}. It leads to the positive semi-definite condition for the BdG Hamiltonian matrix $H({\bf k})$, i.e., $H({\bf k})\geq0$.

\subsection{Non-Hermitian dynamical matrix and spectral decomposition\label{sec:DM}}
Equipped with the canonical commutation relations (\ref{eq:CCR1}) and (\ref{eq:CCR2}), we obtain the equation of motion of the free-boson system in the momentum space
\begin{equation}
	\partial_t\hat{\Phi}_{\bf k}\left(t\right)=-{\rm i}[\hat{\Phi}_{\bf k}(t),\hat{H}]=-{\rm i}H_\tau({\bf k})\hat{\Phi}_{\bf k}(t),
	\label{eq:EOM}
\end{equation}
where $H_{\tau}({\bf k})=\tau_{3}H({\bf k})$ is the dynamical matrix. From it, one can see that the evolution equation is non-unitary since $H_\tau({\bf k})$ is non-Hermitian. Only in this way it is ensured that the Nambu spinor $\hat{\Phi}_{\bf k}(t)$ fulfills the canonical commutation relations (\ref{eq:CCR1}) and (\ref{eq:CCR2}) arising from the Bose-Einstein statistics.

The spectrum of the BdG Hamiltonian (\ref{eq:H1}) can be obtained from the eigenvalues of the dynamical matrix $H_{\tau}({\bf k})$. Explicitly, the positive eigenvalues of $H_{\tau}({\bf k})$ provide the quasiparticle excitation spectrum for the free-boson system in the thermodynamic-stability regime. To see it, we define a Bogoliubov transformation
\begin{equation}
\hat{\Phi}_{\bf k}=V({\bf k})\left(\begin{array}{c}
		\hat{\beta}_{\bf k}\\
		\hat{\beta}^{\dagger T}_{-{\bf k}}
	\end{array}\right),\ \ V({\bf k})=\left(\begin{array}{cc}
		X({\bf k}) & Y^*(-{\bf k})\\ Y({\bf k}) & X^*(-{\bf k})\\
	\end{array}\right),
\label{eq:Trans1}
\end{equation}
where $\hat{\beta}_{\bf k}=(\hat{\beta}_{{\bf k}1},\hat{\beta}_{{\bf k}2},\ldots,\hat{\beta}_{{\bf k}\tilde{N}})^T$ is the array of the annihilation operators. In this work, we mainly focus on positive definite $H({\bf k})$, and the Hamiltonian with zero-energy excitation mode can be regarded as a special limit case. Thanks to the Williamson theorem~\citep{Williamson1936AJM, Simon1999JMP}, the matrix $R({\bf k})$ can always be symplectically diagonalized in the quadrature representation, i.e., $R'({\bf k})=J^T({\bf k})R({\bf k})J({\bf k})$, where $R'({\bf k})$ is a real and diagonal matrix, and $J({\bf k})$ satisfies the symplectic condition $J^T({\bf k}){\rm i}\tau_{2}J({\bf k})={\rm i}\tau_{2}$. Hence, one can also perform the Bogoliubov transformation (\ref{eq:Trans1}) to achieve
\begin{align}
	\hat{H}=&\frac{1}{2}\sum_{\bf k}\left(\hat{\beta}^\dagger_{\bf k}\ \hat{\beta}^T_{-{\bf k}}\right) \tau_3\Lambda({\bf k}) \left(\begin{array}{c}
		\hat{\beta}_{\bf k}\\
		\hat{\beta}^{\dagger T}_{-{\bf k}}
	\end{array}\right)-\frac{1}{2}{\rm Tr}K \notag\\
=&\sum_{\bf k} \hat{\beta}^\dagger_{\bf k} E({\bf k}) \hat{\beta}_{\bf k},
	\label{eq:Diag1}
\end{align}
where $\Lambda({\bf k})= {\rm diag} (E({\bf k}), -E(-{\bf k}))$, and $E({\bf k})\geq 0$ is a $\tilde{N}$-dimensional diagonal matrix and describes the Bogoliubov excitation spectrum of the bosonic system. Since the spectral decomposition does not change the nature of bosons, the bosonic commutation relation retains, i.e., $[\hat{\beta}_{{\bf k}i},  \hat{\beta}_{{\bf k}'j}^\dagger] =\delta_{{\bf k}{\bf k}'}\delta_{ij}$ with $i,j=1,2,\ldots,\tilde{N}$. Based on this, the transformation matrix $V({\bf k})$ must satisfy the pseudo-unitary structure
\begin{equation}
	V^\dagger({\bf k}) \tau_3V({\bf k}) =V({\bf k})\tau_3V^\dagger({\bf k}) =\tau_3,
	\label{eq:PseudoTrans}
\end{equation}
where $V(\bf k)$ belongs to the pseudo-unitary group ${\rm U}(\tilde{N},\tilde{N})$. Equipped with Eqs.\,(\ref{eq:Trans1})-(\ref{eq:PseudoTrans}), we obtain the eigenfunction 
\begin{equation}
	H_{\tau}({\bf k})V({\bf k})=V({\bf k})\Lambda({\bf k}).
	\label{eq:Diag2}
\end{equation}
Equation (\ref{eq:Diag2}) shows us that the spectrum of the system is given by the eigenvalues of the non-Hermitian $H_{\tau}({\bf k})$ and its right eigenstates also determine the diagonalizing matrix $V({\bf k})$.

\subsection{Reduction of the bosonic BdG Hamiltonian\label{sec:Reduction}}

Here we review that the bosonic BdG Hamiltonian with particle-number-nonconserving terms ($M({\bf k})\neq 0$) can be transformed to a particle-number-conserving Hamiltonian, i.e., a single-particle one, if $H({\bf k})$ is positive definite~\citep{Zhou2020JPA}. Due to the pseudo-unitary structure (\ref{eq:PseudoTrans}) and $\tau_1 V^*({\bf k}) \tau_1=V({\bf k})$, there is an essential observation
\begin{align}
\!\! V({\bf k})=&e^{W({\bf k})}\left(\begin{array}{cc}
		U({\bf k}) & \\
		& U^*(-{\bf k})\\
\end{array}\right),\label{eq:Decomposition1} \\  
\!\! W({\bf k})=&\frac{1}{2}\log \left[V({\bf k})V^\dagger({\bf k})\right]=\left(\begin{array}{cc}
& \bar{W}({\bf k})\\ \bar{W}^*(-{\bf k}) & \\
\end{array}\right),\label{eq:K_solution}
\end{align}
where $U({\bf k})$ is unitary and $\bar{W}({\bf k})=\bar{W}^T(-{\bf k})$. Here $e^{W({\bf k})}\in {\rm SU}(\tilde{N},\tilde{N})$ satisfies $e^{W({\bf k})} \tau_{3}e^{W({\bf k})}=\tau_{3}$ and $\det e^{W({\bf k})}=e^{{\rm Tr}W({\bf k})} =1$. Then the diagonalization (\ref{eq:Diag2}) can be rewritten as
\begin{align}
H_{\tau}({\bf k})e^{W({\bf k})}=&e^{W({\bf k})}\left( \begin{array}{cc}
	\tilde{K}({\bf k})& \\ & \tilde{K}^T(-{\bf k}) \\
\end{array}\right),\label{eq:Decomposition2}
\end{align}
where $\tilde{K}({\bf k}) =U({\bf k})E({\bf k})U^\dagger({\bf k})$ is positive semi-definite due to $E({\bf k})\geq 0$. Thus, the non-Hermitian dynamical matrix $H_\tau({\bf k})$ has been transformed to a block-diagonal Hermitian one. Substituting Eqs.\,(\ref{eq:Trans1}) and (\ref{eq:Decomposition1}) back into the BdG Hamiltonian, we arrive at a single-particle Hamiltonian,
\begin{align}
\hat{H}=&\frac{1}{2}\sum_{\bf k}\hat{\Phi}^\dagger_{\bf k} e^{-W({\bf k})}\left( \begin{array}{cc}
	\tilde{K}({\bf k})& \\ & \tilde{K}^T(-{\bf k}) \\
\end{array}\right)e^{-W({\bf k})}\hat{\Phi}_{\bf k}\notag\\
&-\frac{1}{2}{\rm Tr}K \notag \\
=&\frac{1}{2}\sum_{\bf k}\tilde{\Phi}^\dagger_{\bf k} \left( \begin{array}{cc}
	\tilde{K}({\bf k})& \\ & \tilde{K}^T(-{\bf k}) \\
\end{array}\right)\tilde{\Phi}_{\bf k}-\frac{1}{2}{\rm Tr}K \notag\\
=&\sum_{\bf k}\tilde{\beta}^\dagger_{\bf k} \tilde{K}({\bf k})\tilde{\beta}_{{\bf k}},
\label{eq:TransH}
\end{align}
where
\begin{equation}
\tilde{\Phi}_{\bf k}=e^{-W({\bf k})}\hat{\Phi}_{\bf k}=\left(\begin{array}{c}
	\tilde{\beta}_{\bf k}\\
	\tilde{\beta}^{\dagger T}_{-{\bf k}}
\end{array}\right),
\label{eq:Sqz}
\end{equation}
with $\tilde{\beta}_{\bf k}=\left(
\tilde{\beta}_{{\bf k}1}, \tilde{\beta}_{{\bf k}2}, \ldots,\tilde{\beta}_{{\bf k}\tilde{N}} \right)^T$ satisfies the commutation relations (\ref{eq:CCR1}) and (\ref{eq:CCR2}). Here $\tilde{K}({\bf k})$ stands for the Bloch Hamiltonian of the system in the quasiparticle basis. Note that Eq.\,(\ref{eq:Sqz}) is called squeezing transformation in quantum optics~\citep{Wan2021PRA}. The ground state of the system is no longer the bosonic vacuum state, but a squeezed vacuum state,
$$\vert{\rm GS}\rangle=\prod_{\bf k}\exp\left(\frac{1}{2}\hat{\Phi}^\dagger_{\bf k}\tau_{3}W({\bf k})\hat{\Phi}_{\bf k}\right)\vert0\rangle,$$
where $\vert0\rangle$ denotes the bosonic vacuum state.

In short, Eq.\,(\ref{eq:TransH}) shows us that the BdG Hamiltonian with particle-number-nonconserving terms ($M({\bf k})\neq 0$) can be reduced to a {\it particle-number-conserving} (single-particle) one for the free-boson system if $H({\bf k})$ is positive definite. This result even holds when the system is dynamically stable~\citep{Flynn2020EPL}. The relevant details of dynamical stability for free-boson system (e.g., connection between thermodynamic and dynamical stability) are provided in Appendix~\ref{sec:DS}. This reduction inherently arises from the bosonic nature as the mapping $e^{W({\bf k})}$ adopted here is a representation transformation.

\section{Hidden sublattice symmetry and enriched topological classification \label{sec:Hidden_Sym_Top}}
\subsection{Topological threefold way \label{sec:TopClass}}
\begin{table}
	\begin{centering}
		\caption{Topological classification for $d$-dimensional ($d$D) fully gapped free-boson systems. Class A, AIII, AI and AII are the symmetry classes for the single-particle Hamiltonian with no symmetry, sublattice symmetry (${\cal S}$), time-reversal symmetry (${\tilde{\cal{T}}\tilde{\cal {T}}}^{*}=+I$) and time-reversal symmetry (${\tilde{\cal{T}}} {\tilde{\cal{T}}}^{*}=-I$), respectively, in the framework of the AZ classification. The second column denotes the classifying space. The entries $\mathbb{Z}$ ($2\mathbb{Z}$), $\mathbb{Z}_{2}$ represent the types of topological bosonic Bogoliubov excitation, while $0$ indicates the absence of nontrivial topological bosonic Bogoliubov excitation.} \label{tab:1}
		\medskip{}
		\begin{tabular*}{8cm}{@{\extracolsep{\fill}}cccccccccc}
			\toprule[.5pt]\specialrule{0em}{.5pt}{.5pt}
			\toprule[.5pt]\specialrule{0em}{1.5pt}{2.5pt}
			\multirow{2}{*}{AZ class} & & \multicolumn{8}{c}{$d$} \\
			\cmidrule[.4pt]{3-10} \cmidrule{4-10} \cmidrule{5-10} \cmidrule{6-10} \cmidrule{7-10} \cmidrule{8-10} \cmidrule{9-10} \cmidrule{10-10}\specialrule{.0pt}{1.5pt}{2.5pt}
			& & $0$ & $1$ & $2$ & $3$ & $4$ & $5$ & $6$ & $7$ \\
			\specialrule{.4pt}{1.5pt}{2.5pt}
			A & ${\cal C}_0$ & $\mathbb{Z}$ & $0$ & $\mathbb{Z}$ & $0$ & $\mathbb{Z}$ & $0$ & $\mathbb{Z}$ & $0$  \\
			\specialrule{.0pt}{1.5pt}{2.5pt}
			AIII & ${\cal C}_1$ & $0$ & $\mathbb{Z}$ & $0$ & $\mathbb{Z}$ & $0$ & $\mathbb{Z}$ & $0$ & $\mathbb{Z}$\\
			\specialrule{.4pt}{1.5pt}{2.5pt} 
			AI & ${\cal R}_0$ & $\mathbb{Z}$ & $0$ & $0$ & $0$ & $2\mathbb{Z}$ & $0$ & $\mathbb{Z}_{2}$ & $\mathbb{Z}_{2}$\\
			\specialrule{.0pt}{1.5pt}{2.5pt}
			AII & ${\cal R}_4$ & $2\mathbb{Z}$ & $0$ & $\mathbb{Z}_{2}$ & $\mathbb{Z}_{2}$ & $\mathbb{Z}$ & $0$ & $0$ & $0$\\
			\specialrule{0em}{1.5pt}{0pt}
			\bottomrule[.5pt]\specialrule{0em}{.5pt}{.5pt}
			\bottomrule[.5pt]
		\end{tabular*}
	\end{centering}
\end{table}
The AZ classification concerns the symmetry properties and topological features of Hamiltonians at the single-particle level for free fermions in terms of the standard time-reversal, particle-hole and chiral symmetries~\citep{Kitaev2009AIP, Ryu2010NJP, Chiu2016RevModPhys}. Inherited from the framework of the AZ classification, here we will show how to rigorously and quickly achieve the topological threefold way (i.e., class A, AI and AII) for the fully gapped free-boson system. Here we adopt the notion of symmetry introduced in the review~\citep{Chiu2016RevModPhys} (more details of the symmetry seen in Appendix~\ref{sec:Sym_Def}). Nevertheless, it is worth noting that the particle-hole and chiral symmetries sometimes are interpreted as “constraints” because they arise naturally for free fermions as “descendants” of special many-body symmetries and lack of physical interpretations~\citep{Xu2020PRB}.

Now, we review the time-reversal, particle-hole and chiral symmetries for the free-boson system, respectively, defined by~\citep{Zhou2020JPA} (More details seen in Appendix~\ref{sec:Sym_Def})
\begin{align}
{\cal{T}}H_{\tau}^{*}(-{\bf k}){\cal{T}}^{-1}=H_{\tau}({\bf k}),&\ \ \ [{\cal{T}},\tau_{3}]=0,\label{eq:TRS} \\
{\cal{C}}H_{\tau}^{*}(-{\bf k}){\cal{C}}^{-1}=-H_{\tau}({\bf k}),&\ \ \ \{{\cal{C}},\tau_{3}\}=0,\label{eq:PHS} \\
\Gamma H_{\tau}({\bf k})\Gamma^{-1}=-H_{\tau}({\bf k}),&\ \ \ \{\Gamma,\tau_{3}\}=0,\label{eq:CS}
\end{align}
where the unitary matrices ${\cal T}, {\cal C}, {\Gamma}$ denote time reversal, particle hole and chiral, respectively, and ${\Gamma}$ is the combination of ${\cal T}$ and ${\cal C}$, i.e., ${\Gamma}={\cal{T}}{\cal C}^*$. The time-reversal and particle-hole symmetries correspond to antiunitary transformations in the Krein space, while the last one corresponds to a unitary transformation. The commutation of ${\cal T}$ and anti-commutation of ${\cal C},\Gamma$ with $\tau_3$ originate from the necessity of the symmetry-preserving bosonic nature. Notably, the bosonic BdG Hamiltonian intrinsically has the particle-hole symmetry with ${\cal C}=\tau_1$. Thus, ${\cal T}$ is block diagonal, i.e., ${\cal T}={\tilde{\cal{T}}}\oplus\tilde{\cal{T}}^*$, and $\Gamma={\cal T}\tau_1$ is block off-diagonal.

Here we argue that in terms of these three symmetries, the topological classification for fully gapped free-boson systems is identical to that of the single-particle systems with Hamiltonian matrix $\tilde{K}({\bf k})$. To verify it, we can utilize a useful lemma:

{\it Lemma 1.} The mapping $e^{W({\bf k})}$ preserves those symmetries that commute or anti-commute with $\tau_{3}$ if the bosonic BdG Hamiltonian is dynamically stable.

The proof is given in Appendix~\ref{sec:Proof}. Here we prove this lemma for the positive-definite case. Suppose $H({\bf k})$ is positive definite, $H({\bf k})>0$, thus, ${W({\bf k})}$ is uniquely determined by 
\begin{equation}
e^{2W}=H^{-1/2}\left(H^{1/2}\tau_3H\tau_3H^{1/2}\right)^{1/2}H^{-1/2}.
\label{eq:WSol}
\end{equation}
It is obtained from the relation $e^{2W}He^{2W}=\tau_3 H\tau_3$. We temporarily suppress the ${\bf k}$-dependence of $H$ and $W$ for brevity. Considering a linear or antilinear internal symmetry $OH_{\tau}(\epsilon_O{\bf k})O^{-1}=\eta_{O} H_{\tau}({\bf k})$, where the unitary or antiunitary matrix $O$ obeys $O\tau_{3}=\eta_{O}\tau_{3}O$, and $\eta_{O}, \epsilon_{O}=\pm$, one can easily find that $Oe^{2W(\epsilon_O{\bf k})}O^{-1}=e^{2W({\bf k})}$, or
\begin{equation}
OW(\epsilon_{O}{\bf k})O^{-1}=W({\bf k}).
\end{equation}
Note that $\epsilon_{O}=-$ is responsible for the symmetry case with an antiunitary transformation. Then, the block-diagonal Hermitian matrix $H'_{\tau}({\bf k})=\tilde{K}({\bf k})\oplus-\tilde{K}^*(-{\bf k})$ satisfies
\begin{equation}
OH'_{\tau}(\epsilon_{O}{\bf k})O^{-1}=\eta_{O}H'_{\tau}({\bf k}).
\end{equation}
Thus, Lemma 1 is true for the positive-definite case.

Suppose there is a band gap (i.e., topological obstruction) in the eigenvalues $\Lambda({\bf k})$ for all ${\bf k}$, and $H_{\tau}({\bf k})$ and $e^{W({\bf k})}$ are continuous with respect to ${\bf k}$~\citep{Continuity}. The dynamical matrix $H_{\tau}({\bf k})$ can be continuously deformed to $H'_{\tau}({\bf k})$ while preserving the gap and the three standard symmetries with the virtue of Lemma 1, e.g.~\citep{Zhou2020JPA},
\begin{equation}
H_{\tau}({\bf k};\lambda)=e^{(1-\lambda)W({\bf k})}H'_{\tau}({\bf k})e^{-(1-\lambda)W({\bf k})}, 
\label{eq:Deform}
\end{equation}
with $\lambda\in[0,1]$. This mapping satisfies $H_\tau({\bf k};\lambda=0)=H_{\tau}({\bf k})$ and $H_{\tau}({\bf k};\lambda=1)=H'_{\tau}({\bf k})$ and preserves the three symmetries. As the mapping $e^{W({\bf k})}$ is intrinsically a Bogoliubov (squeezing) transformation, $H_{\tau}({\bf k})$ and $H'_{\tau}({\bf k})$ share the same eigenvalues and, hence, the band gap. Here we prove that the continuous mapping adopted from Ref.\,\citep{Zhou2020JPA} even holds in the dynamical-stability regime. We also note that our recipe is also similar to the one derived via the polar decomposition in Ref.~\citep{Gong2022arXiv}. Therefore, the topological classification of the dynamical matrix $H_{\tau}({\bf k})$ is identical to topologically classifying all the possible blocks of $H'_{\tau}({\bf k})$, i.e., $\tilde{K}({\bf k})$, in terms of the time-reversal, particle-hole and chiral symmetries. It is worth noting that although this mapping does not preserve all symmetries, e.g., total-number symmetry, here only these three symmetries are concerned for topologically classifying the free-boson system.

Now we focus on the block $\tilde{K}({\bf k})$ and classify its ensemble in terms of the three symmetries. The block $\tilde{K}({\bf k})$ describes a single-particle system in ${\bf k}$ space and has no particle-hole-mixing term (i.e., particle-number-nonconserving term). It means the total-number symmetry recovers, but the particle-hole and chiral symmetries are dismissed~\citep{Ref1}. As a consequence, we just keep the time-reversal symmetry in consideration, which is then given by
\begin{equation}
{\tilde{\cal{T}}} \tilde{K}^*(-{\bf k}){\tilde{\cal{T}}}^{-1}=\tilde{K}({\bf k}), \ \ \ {\tilde{\cal{T}}}{\tilde{\cal{T}}}^*=\pm I.
\label{eq:TRS1}
\end{equation}
Armed with the framework of the AZ classification, we obtain class A (absence of time-reversal symmetry), AI (${\tilde{\cal{T}}}{\tilde{\cal{T}}}^*=+I$) and AII (${\tilde{\cal{T}}} {\tilde{\cal{T}}}^*=-I$) for the free-boson system, which are also known as the standard threefold way~\citep{Dyson1962JMP, Zirnbauer2015}.

So far, we have discussed the diagonalizable case of the dynamical matrix $H_{\tau}({\bf k})$. In fact, one may encounter the non-diagonalization of $H_{\tau}({\bf k})$ when the bosonic BdG Hamiltonian matrix is positive semi-definite, $H({\bf k})\geq 0$~\citep{Flynn2020NJP, Exception}. Such a case describes the emergence of the exceptional point at zero energy, at which energies and eigenstates both coalesce~\citep{Heiss2012JPA}. In fact, it can be addressed safely by adding an infinitesimal onsite energy $\Delta I$ ($\Delta\rightarrow 0^{+}$) to $H({\bf k})$ since this treatment would not influence the topological bosonic Bogoliubov excitation of the system at zero energy (See Appendix~\ref{sec:No_go} for details).

\subsection{Hidden sublattice symmetry and enriched topological classification of bosonic Bogoliubov excitations}

For free-fermion systems, especially superconductors (with pairing terms), zero energy is usually assumed as the band gap, and the ground state can have a nontrivial topology and host in-gap edge (surface) states because states below the gap are fully occupied at zero temperature. In contrast, the bosonic ground state must have a trivial topology in the free-boson system~\citep{Lu2018arXiv}, which has been summarized by a no-go theorem~\citep{Xu2020PRB}. But one can still study the bosonic Bogoliubov excitation with finite energy since it could be topologically nontrivial as predicted by the threefold way, i.e., class A, AI and AII in Table\,\ref{tab:1}. However, the threefold way still shows topological triviality of bosonic Bogoliubov excitation for the free-boson system in one dimension. It makes us wonder if there are any other internal symmetries to support the 1D topological bosonic excitation.

Here we will show that the free-boson system can have an additional internal symmetry, i.e., sublattice symmetry, which can supply nontrivial topological excitations in {\it odd} dimension.

We first focus on the single-particle Hamiltonian matrix $\tilde{K}({\bf k})$ ($\tilde{K}({\bf k})\geq0$) as the dynamically stable BdG Hamiltonian can be reduced to a particle-number-conserving one based on the aforementioned mapping, or explicitly, pseudo-unitary transformation $e^{W({\bf k})}$. Sublattice symmetry for the system in the quasiparticle basis is defined as
\begin{equation}
\tilde{{\cal S}} h({\bf k}) \tilde{{\cal S}}^{-1}=-h({\bf k}), \ \ \ \tilde{{\cal S}}^2=I,
\label{eq:SLS1}
\end{equation}
where $h({\bf k})=\tilde{K}({\bf k})-\epsilon I$ denotes the {\it traceless} part of the single-particle Hamiltonian matrix and $\epsilon={\rm Tr}\tilde{K}/\tilde{N}$ is an onsite energy independent of ${\bf k}$, and $\tilde{{\cal S}}$ is a unitary and Hermitian matrix, $\tilde{{\cal S}}=\tilde{{\cal S}}^\dagger =\tilde{{\cal S}}^{-1}$. Note that the onsite energy $\epsilon$ is a necessity to guarantee the existence of ground state for the bosonic system, as it keeps the spectrum non-negative. This symmetry gives rise to a symmetric spectrum of the single-particle system with respect to the finite energy $\epsilon$: if $\vert{u}_{j}({\bf k})\rangle$ is an eigenstate of $\tilde{K}({\bf k})$ with energy $E_{j}({\bf k})\geq0$, then $\tilde{S}\vert{u_{j}({\bf k})}\rangle$ is also an eigenstate but with energy $(2\epsilon-E_{j}({\bf k}))\geq0$. One can see that $\tilde{\cal S}$ is a unitary transformation on the underlying Hilbert space for the single-particle system, $\tilde{\cal S}:\mathbb{H}\rightarrow\mathbb{H}$. In the diagonal form of $\tilde{\cal{S}}$, $h({\bf k})$ is block off-diagonal and can be written as
\begin{equation}
h({\bf k})=\left(\begin{array}{cc}
&D({\bf k})\\ D^\dagger({\bf k}) &
\end{array}\right),
\end{equation}
where $D({\bf k})$ is a $\tilde{N}_{A}\times \tilde{N}_{B}$ rectangular matrix and $\tilde{N}_{A/B}$ ($\tilde{N}_{A}+\tilde{N}_{B}=\tilde{N}$) is the number of sites on sublattice $A/B$. As an example, sublattice symmetry appears in bipartite lattices where particle hopping only connects sites on different sublattices, such as the Su-Schrieffer-Heeger (SSH) model with nonzero onsite energy.

In the original basis, this symmetry can be identified by the pair of non-negative eigenvalues of $H_{\tau}({\bf k})$, i.e., $(E_{j}({\bf k}),\,\epsilon-E_{j}({\bf k}))\geq0$, with the corresponding eigenstates $(\vert\phi_{j}({\bf k})\rangle,\,{\cal S}\vert\phi_{j}({\bf k})\rangle)$, respectively, where $\vert\phi_{j}({\bf k})\rangle=e^{W({\bf k})} (\vert u_{j}({\bf k})\rangle^T, 0 )^T$ and ${\cal S}=\tilde{\cal S}\oplus\tilde{S}^*$. Here we have assumed that the mapping also preserves sublattice symmetry, ${\cal S}e^{W({\bf k})}=e^{W({\bf k})}{\cal S}$. Notice that these pairs have the positive square norm, i.e., $\langle \phi_{j}({\bf k})\vert\tau_{3}\vert\phi_{j}({\bf k})\rangle=\langle \phi_{j}({\bf k})\vert{\cal S}^{\dagger}\tau_{3}{\cal S}\vert\phi_{j}({\bf k})\rangle>0$. Such a symmetry acting on the subspaces of the Krein space ($\mathbb{J}_{\pm}$) can be immediately applied to invert the dynamics of this free-boson system ${\cal U}(t)=\exp(-{\rm i}H_{\tau}({\bf k})t)$:
\begin{equation}
{\cal U}(-t)=\exp{\left(2{\rm i}t\epsilon\tau_{3}e^{-2W({\bf k})}\right)}{\cal S}{\cal U}(t){\cal S}^{-1}.
\end{equation}
The additional term $\epsilon\tau_3e^{-2W({\bf k})}$ corresponds to a particle-number-nonconserving one in the bosonic BdG Hamiltonian. But, it describes a uniform onsite potential with finite energy $\epsilon$ when the Hamiltonian is transformed into the quasiparticle basis. More details of sublattice symmetry on the many-particle space are provided in Appendix~\ref{sec:SLS_MP}.

In terms of a band gap at energy $\epsilon$, the positive-definite matrix $e^{\pm W({\bf k})}$  preserving sublattice symmetry considered here can be continuously deformed to an identity matrix in the dynamical matrix $H_{\tau}({\bf k})$, as shown in Eq.\,(\ref{eq:Deform}). Therefore, the topological classification of these BdG Hamiltonians is equivalent to topologically classify the family of the single-particle ones with sublattice symmetry. In the framework of AZ classification, these single-particle Hamiltonians are actually classified to class AIII as the sublattice symmetry is mathematically similar to  chiral symmetry in free-fermion systems. In this class, there may exist topological bosonic Bogoliubov excitations supported by sublattice symmetry in odd dimension, as shown in Table\,\ref{tab:1}.

Before proceeding, we provide some remarks on the proposed sublattice symmetry (\ref{eq:SLS1}). Firstly, this symmetry is usually hidden in terms of the original dynamical matrix $H_{\tau}({\bf k})$, since the implicit term $\epsilon\tau_{3}e^{-2W({\bf k})}$ is not easily distinguished from the dynamical matrix $H_{\tau}({\bf k})$.

Secondly, sublattice symmetry is not physically identical to the chiral symmetry at the single-particle level in the free-boson system even if Eqs.\,(\ref{eq:CS}) and (\ref{eq:SLS1}) share the same mathematical form. This is because the chiral operation $\Gamma$ mixes the particles and holes while the sublattice operation ${\cal S}$ does not, i.e., $\{\Gamma,\tau_{3}\}=0$ and $[{\cal S},\tau_{3}]=0$. It also implies that the combination of time-reversal symmetry and sublattice symmetry is not particle-hole symmetry.

Thirdly, the sublattice symmetry is actually ubiquitous in bosonic systems. Explicitly, bipartite lattices where particle only tunnels to different sublattices can be realized in many platforms, such as the photonic superlattice\,\citep{Malkova2009OL}, bosonic cold atoms loaded in a double-well potential~\citep{Lee2007PRL}, plasmonic waveguide arrays~\citep{Bleckmann2017PRB, Cherpakova2018LSA}, etc.

\subsection{A construction of the bosonic BdG Hamiltonian with sublattice symmetry \label{sec:Construction}}
In the above, we have shown that the free-boson system could possess a sublattice symmetry through an implicit single-particle Hamiltonian matrix $h({\bf k})$. But it drives us wonder what kind of the bosonic BdG Hamiltonian matrix $H({\bf k})$ should be. Here we provide a possible construction of the bosonic BdG Hamiltonian which has the sublattice symmetry.

Suppose that the matrix $K({\bf k})=\varepsilon I+h({\bf k})$ ($\varepsilon>0$), and $h({\bf k})$ respects the time-reversal symmetry (\ref{eq:TRS1}) with ${\tilde{\cal{T}}}=I$ and sublattice symmetry (\ref{eq:SLS1}). Notice that $K({\bf k})$ corresponds to the single-particle part in the free-boson system (\ref{eq:H1}). For the particle-number-nonconserving term, we also assume the matrix $M({\bf k})=\xi {\cal S}$, with $\xi\in\mathbb{C}$ and the sublattice operation ${\cal S}={\cal S}^T$ being symmetric. Then the associated bosonic BdG Hamiltonian matrix can be rewritten as 
\begin{equation}
H({\bf k})=\left(\begin{array}{cc}
\varepsilon I+h({\bf k}) & \xi {\cal S}\\
\xi^* {\cal S} & \varepsilon I+h({\bf k})\\
\end{array}\right).
\label{eq:H3}
\end{equation}
Now let us show that the free-boson system with Eq.\,(\ref{eq:H3}) respects the sublattice symmetry. Following the block-diagonalization treatment discussed in Sec.\,\ref{sec:Reduction}, we perform a pseudo-unitary transformation $e^{W({\bf k})}\in{\rm SU}(N,N)$ with 
\begin{equation}
W({\bf k})=\left(\begin{array}{cc}
	&\Xi {\cal S}\\
	\Xi^* {\cal S} & \\
\end{array}\right)
\label{eq:W2}
\end{equation}
and $\Xi=e^{{\rm i}\phi}|\Xi|\in \mathbb{C}$ such that the dynamical matrix $H_{\tau}({\bf k})$ can be block-diagonalized, i.e., 
\begin{equation}
H_{\tau}({\bf k})e^{W({\bf k})}=e^{W({\bf k})}\left(\begin{array}{cc}
\tilde{K}({\bf k}) & \\
 & -\tilde{K}^T(-{\bf k})\\
\end{array}\right).
\label{eq:Block_Diag}
\end{equation}
After some algebra, the solutions of $|\Xi|$ and $\phi$ are obtained by 
\begin{equation}
|\Xi|=\frac{1}{2}\log \sqrt{\frac{\varepsilon-|\xi|}{\varepsilon+|\xi|}}, \ \ \ \phi=\arg(\xi).
\end{equation}
Thus, the single-particle Hamiltonian matrix in the new quasiparticle basis is given by
\begin{equation}
\tilde{K}({\bf k})={\epsilon}I+h({\bf k}),
\end{equation}
where $\epsilon=\sqrt{\varepsilon^2-|\xi|^2}$. It can been seen that the system has the sublattice symmetry (\ref{eq:SLS1}). Therefore, we have proved that the category of free-boson systems with Eq.\,(\ref{eq:H3}) respects the sublattice symmetry.

\subsection{Comparison to the existing literature}
Recently, the topological classification for free-boson systems has been demonstrated in Refs.\,\citep{Lu2018arXiv, Kawabata2019PRX, Lein2019PRB, Zhou2020JPA, Xu2020PRB, Chaudhary2021PRB}. It is necessary to discuss the relation and crucial distinction between these works and ours.

In Refs.\,\citep{Lu2018arXiv, Kawabata2019PRX, Zhou2020JPA}, the standard K-theory approach~\citep{Kitaev2009AIP} is exploited to achieve the topological classes, i.e., class A, AI and AII, for free-boson systems in the thermodynamic-stability regime, as presented in Table\,\ref{tab:1}. Such an approach is powerful, but complex and hard to follow. Apart from this, the insight that the particle-hole symmetry is intrinsically a constraint rather than a symmetry in Refs.\,\citep{DeNittis2019, Lein2019PRB} is deep, but also incomprehensible. In contrast, our method of the Hamiltonian reduction shown in Sec.\,\ref{sec:TopClass} provides a easy and rigorous shortcut to obtain the topological classification based on the framework of AZ classification. We note that the constraint mentioned above does not carry any symmetry properties of Hamiltonians at the single-particle level concerned in the AZ classification, and this notion is totally distinguished from the particle-hole symmetry (sometimes dubbed particle-hole constraint~\citep{Xu2020PRB}) of free-fermionic systems.

An alternative approach is based on the squaring map from a fermionic BdG Hamiltonian matrix $H_f({\bf k})$ to a bosonic BdG one~\citep{Xu2020PRB}. In this case, the particle-hole symmetry and chiral symmetry disappear and the constructed bosonic BdG Hamiltonian $H_f^2({\bf k})$ is classified into the standard topological threefold way. However, as discussed in Ref.\,\citep{Chaudhary2021PRB}, the squaring map $H^2_{f}({\bf k})$ is not surjective onto the collection of any BdG Hamiltonian matrices $H({\bf k})$. In other words, $H_f^2({\bf k})$ cannot represent any bosonic BdG Hamiltonian matrices. As a contrast, we have shown that a particle-number-nonconserving bosonic BdG Hamiltonian {\it one-to-one} corresponds to a particle-number-conserving one through a special pseudo-unitary transformation $e^{W({\bf k})}$ if the free-boson system is dynamically stable~\citep{Flynn2020EPL}.

Moreover, Ref.\,\citep{Chaudhary2021PRB} has proposed an adiabatic mapping (i.e., continuous deformation) from a particle-number-nonconserving bosonic BdG Hamiltonian to a particle-number-conserving one, which preserves symmetries commuting with $\tau_{3}$. Actually, their mapping is topologically equivalent to ours, although it does not keep the excitation spectrum of the system. One can obtain their proposed mapping by the replacement $e^{\pm(1-\lambda)W({\bf k})}\rightarrow [\cosh W({\bf k})\pm(1-\lambda)\sinh W({\bf k})]$ in Eq.\,(\ref{eq:Deform}), with $\lambda\in[0,1]$, and thus arrive at a particle-number-conserving Hamiltonian. The replacement does not close the band gap and preserves the referred symmetries due to the positive definiteness of $e^{W({\bf k})}$ and $\cosh W({\bf k})$. Notably, we have also proved that the proposed mapping $e^{W({\bf k})}$ preserves symmetries that {\it anti-commute} with $\tau_{3}$. Besides, we have provided a physical picture of the reduction of the dynamically stable BdG Hamiltonian: the free-boson system with particle-number-nonconserving terms is actually a single-particle one with the squeezed vacuum state $\vert{\rm GS}\rangle$ in the quasiparticle basis.

In addition to these differences, we also find the internal sublattice symmetry (\ref{eq:SLS1}) in the free-boson system. Systems with this symmetry are classified to an additional topological class, i.e., class AIII, and can have a $\mathbb{Z}$-type topological bosonic Bogoliubov excitation in odd spatial dimension shown in Table\,\ref{tab:1}. It is worth noting that spatial symmetry can also support topological bosonic excitations in 1D systems. For example, high-lying topological band supported by inversion symmetry has been reported in 1D optical lattice loaded with bosonic cold atoms~\citep{Engelhardt2015PhysRevA}. Actually, additional spatial symmetry can modify the topological classification~\citep{Fu2011PRL}. However, the purpose of this work is to find topological bosonic Bogoliubov excitations in one dimension guaranteed by additional {\it internal} symmetries. Our obtained topological classification enriches the current threefold way and opens a new avenue to study topological physics of the free-boson system.

\section{Topological bosonic Bogoliubov excitation in one dimension \label{sec:ExI}}

\subsection{Prototypical model}

\begin{figure}
	\begin{centering}
		\includegraphics[width=8.5cm]{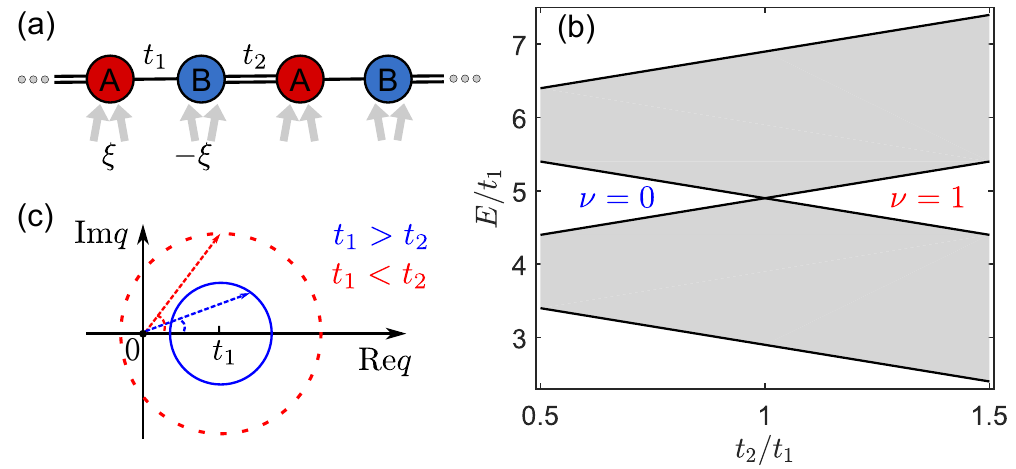}
	\end{centering}
	\caption{(a) The prototypical model whose unit cell is formed by the sublattices $A$ and $B$. The shaded arrows denote the bi-particle processes at sublattice $A$ with $\xi=|\xi|e^{{\rm i}\phi}$ and $B$ with $-\xi$. (b) The Bloch excitation bands of the BdG Hamiltonian matrix $H_{\rm I}(k)$ (shaded areas) characterized by the winding number $\nu$ as a function of $t_2$ for $\mu=5t_1$ and $\xi=t_1$. (c) The circles defined by $q(k)$ in the trivial (solid-blue circle) and topological (dashed-red circle) invariants. For $t_1<t_2$, the dashed-red circle encloses the origin once while the winding number is vanishing for $t_1>t_2$.}
	\label{fig:1}
\end{figure}

Here we resort to a 1D prototypical model to illustrate the nontrivial topological bosonic Bogoliubov excitation with the hidden sublattice symmetry. Specifically, suppose that there is a dimer chain formed by bosonic modes and each mode is coupled to nearest-neighbor sites, as shown in Fig.\,\ref{fig:1}(a). The associated BdG Hamiltonian of such a system $\hat{H}_{\rm I}$ is divided into two parts: the single-particle part $\hat{H}_{\rm I}^{(1)}$ and particle-number-nonconserving term $\hat{H}_{\rm I}^{(2)}$. The former part is expressed as 
\begin{align}
\hat{H}_{\rm I}^{(1)}=\sum_{j=1}^{L}&\left[\mu\sum_{s=\pm}\hat{a}^\dagger_{j,s}\hat{a}_{j,s}+(t_1\hat{a}_{j,+}^\dagger\hat{a}_{j,-}\right. \notag\\
&\left.+t_2\hat{a}^\dagger_{j+1,+}\hat{a}_{j,-}+{\rm H.c.})\right],
\end{align}
where $\mu>0$ is the onsite potential and $t_1,t_2>0$ are the intra- and intercell couplings, respectively. Here $\hat{a}_{j,s}$ ($\hat{a}_{j,s}^\dagger$) are the bosonic annihilation (creation) operators, and $j$ denotes the $j$-th unit cell while the subscripts $s=\pm$ label the sublattices $A$ and $B$, respectively. The latter is given by 
\begin{equation}
\hat{H}_{\rm I}^{(2)}=\frac{\xi}{2}\sum_{j=1}^{L}\sum_{s=\pm} s\hat{a}^\dagger_{j,s}\hat{a}_{j,s}^\dagger+{\rm H.c.},
\end{equation}
where $\xi=|\xi|e^{{\rm i}\phi}\in\mathbb{C}$, $|\xi|$ and $\phi$ denote the strength and phase of the particle-number-nonconserving term, respectively. Subject to the translation-invariance and periodic boundary condition, the bosonic BdG Hamiltonian can be transformed into the momentum space, and obtained by $\hat{H}_{\rm I}=\frac{1}{2}\sum_{k}\hat{\Phi}^\dagger_kH_{\rm I}(k)\hat{\Phi}_{k}$, where $\hat{\Phi}_{k}=(\hat{a}_{k+},\hat{a}_{k,-},\hat{a}^\dagger_{-k,+},\hat{a}^\dagger_{-k,-})^T$ is the bosonic Nambu spinor of the chain and $k\in [0,2\pi)$ is the momentum. Note that we have neglected the constant in the BdG Hamiltonian. The Bloch BdG Hamiltonian matrix reads
\begin{align}
H_{\rm I}(k)=&\left(\begin{array}{cc}
H^{(1)}_{\rm I}(k) & H^{(2)}_{\rm I}\label{eq:BlochH1}\\
\left[H^{(2)}_{\rm I}\right]^* & \left[H^{(1)}_{\rm I}(-k)\right]^*\\
\end{array}\right),\\
H^{(1)}_{\rm I}(k)=&\mu I+(t_1+t_2\cos k)\sigma_1 +t_2\sin k\sigma_2,\notag\\
H^{(2)}_{\rm I}=&\xi\sigma_3,\notag
\end{align}
where $\sigma_{1,2,3}$ are the conventional Pauli matrices. 

From Eq.\,(\ref{eq:Diag1}), the bosonic BdG Hamiltonian can be diagonalized via a Bogoliubov transformation, i.e., 
\begin{equation}
	\hat{H}_{\rm I}=\sum_{k}\sum_{s=\pm}E_{s}(k)\hat{\gamma}^\dagger_{k,s} \hat{\gamma}_{k,s},
\end{equation}
where the Bloch excitation spectrum is
\begin{equation}
E_{\pm}(k)=\tilde{\mu}\pm\sqrt{t_1^2+t_2^2+2t_1t_2\cos k},
\end{equation}
with $\tilde{\mu}=\sqrt{\mu^2-|\xi|^2}$. Here the annihilation operators of the quasiparticle are $\hat{\gamma}_{k,s}=\langle v_{s}(k)\vert\tau_{3}\hat{\Phi}_k$ and $\vert v_{s}(k)\rangle$ are the eigenstates of the dynamical matrix $H_{\tau \rm I}(k)=\tau_3H_{\rm I}(k)$. It can be seen that a excitation gap is open at $\tilde{\mu}$ if $t_1\neq t_2$, which is shown in Fig.\,\ref{fig:1}(b). By the way, $\tilde{\mu}\geq t_1+t_2$ is assumed for the thermodynamic stability.

\subsection{Topological invariant and its measurement}

Now let us show that such a 1D free-boson system supplies nontrivial topological invariant. We first show that the single-particle part $\hat{H}^{(1)}_{\rm I}$ in the bosonic BdG Hamiltonian respects the sublattice symmetry in the interaction picture with $\hat{U}_{\rm I}(t)=\prod_{k,s}\exp(-{\rm i}\mu \hat{a}_{k,s}^\dagger \hat{a}_{k,s}t)$. That is, one can obtain Eq.\,(\ref{eq:SLS1}) by the replacements $h(k)\rightarrow h^{(1)}_{\rm I}(k)=[H^{(1)}_{\rm I}(k)-\mu I]$ and ${\cal S}\rightarrow\sigma_3$. It is worthy to mention that the Hamiltonian matrix $h^{(1)}_{\rm I}(k)$ describes the SSH model. It can be readily to verify that the single-particle Hamiltonian respects the time-reversal symmetry 
\[{\tilde{\cal{T}}}[H_{\rm I}^{(1)}(-k)]^*{\tilde{\cal{T}}}^{-1}=H_{\rm I}^{(1)}(k),\ \ {\tilde{\cal{T}}}=I,\] 
and the matrix of the particle-number-nonconserving term is symmetric $(H^{(2)}_{\rm I})^T=H^{(2)}_{\rm I}$. Therefore, following the treatment in Sec.\,\ref{sec:Construction}, we perform the pseudo-unitary transformation $e^W$ with 
\[W=r\left(\begin{array}{cc}
 & e^{{\rm i}\phi}\sigma_3\\
 e^{-{\rm i}\phi}\sigma_3& \\
\end{array}\right),\] 
$r=\log \sqrt{(\mu-|\xi|)/(\mu+|\xi|)}/2$.
Then the bosonic BdG Hamiltonian can be transformed to a single-particle one, i.e.,  
\begin{align}
\hat{H}_{\rm I}=\sum_{k}&\left\{\tilde{\mu} (\hat{\beta}_{k,+}^\dagger \hat{\beta}_{k,+}+ \hat{\beta}_{k,-}^\dagger \hat{\beta}_{k,-})\right. \notag\\
&\left. +[(t_1+t_2e^{-{\rm i}k})\hat{\beta}_{k,+}^\dagger\hat{\beta}_{k,-}+{\rm H.c.}]\right\},
\end{align}
where $\hat{\beta}_{k,\pm}=\cosh r\hat{a}_{k,\pm}\pm e^{{\rm i}\phi}\sinh r\hat{a}_{-k,\pm}^{\dagger}$. The associated Bloch Hamiltonian matrix becomes 
\begin{equation}
\tilde{H}_{\rm I}(k)=\tilde{\mu}I+(t_1+t_2\cos k)\sigma_1+t_2\sin k\sigma_2.
\label{eq:BlochH2}
\end{equation}
Explicitly describing the SSH model, such a Hamiltonian respects the sublattice symmetry and is classified to class AIII. It implies that this model can have nontrivial topological bosonic Bogoliubov excitation. To characterize its topology, we can define the winding number 
\begin{equation}
\nu=\frac{1}{2\pi{\rm i}}\int_{\rm BZ}q^{-1}(k){\rm d}q(k)=\frac{1}{2\pi}\int_0^{2\pi}\partial_{k}\varphi(k){\rm d}k
\label{eq:Winding}
\end{equation}
where $q(k)=t_1+t_2e^{{\rm i}k}$ and $\varphi(k)=\arg(q(k))$. It turns out that the winding number $\nu$ is nontrivial only if $t_1<t_2$ is satisfied. 

Alternatively, the symplectic polarization of the 1D system $P$ is also capable of characterizing the topology of the high-lying excitation. The symplectic polarization, by definition, is given by
\begin{equation}
P=\frac{1}{2\pi}\int_0^{2\pi}{\rm d}k{\cal A}(k), \ \ {\cal A}(k)={\rm i}\langle v_{-}(k)\vert\tau_{3}\vert\partial_k v_{-}(k)\rangle,
\end{equation}
where ${\cal A}(k)$ is the Berry connection and $\vert v_{-}(k)\rangle$ is the lower excitation band of the Bloch BdG Hamiltonian matrix (\ref{eq:BlochH1}). We note that the inner product has the symplectic structure $\tau_{3}$ due to the commutation relation (\ref{eq:CCR2}). In general, the symplectic polarization, which is a geometrical quantity, is not quantized because of the gauge-dependence of the Berry connection ${\cal A}(k)$. In the presence of the sublattice symmetry, the symplectic polarization $P$ is quantized in units of $1/2$ (see the proof in Appendix\,\ref{app:Derivation}). Considering the eigenstate of the lower band $$\vert v_{-}(k)\rangle=\frac{1}{\sqrt{2}}e^{W}\left(\begin{array}{c}
e^{-{\rm i}\varphi(k)}\\ 1\\ 0\\ 0
\end{array}\right),$$
we obtain the symplectic polarization of the system
\begin{equation}
P=\frac{1}{4\pi}\int_0^{2\pi}\partial_{k}\varphi(k){\rm d}k=\frac{\nu}{2}. 
\end{equation}
Eliminating the U(1) gauge redundancy, we have a universal relation between the symplectic polarization and winding number, i.e., $P=\nu/2\mod 1$~\citep{Mondragon-Shem2014PhysRevLett}.

\begin{figure}
	\begin{centering}
		\includegraphics[width=7.5cm]{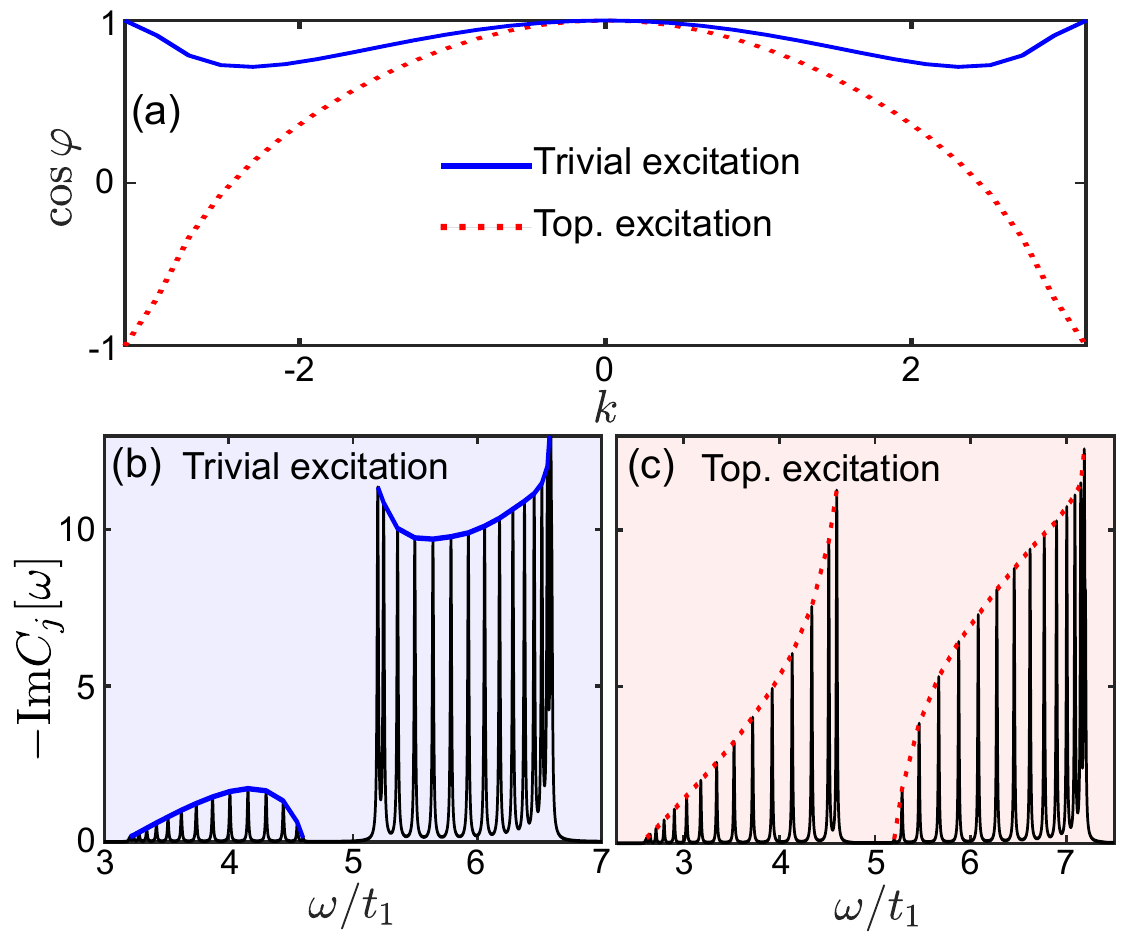}
	\end{centering}
	\caption{(a) $\cos\varphi$ versus $k$. Solid-blue and dashed-red curves represent the trivial and topological excitations, respectively. (b),(c) The imaginary part of the correlation function $-{\rm Im}C_j[\omega]$ versus frequency $\omega$ for the trivial (b) and topological (c) excitations. Here the number of the unit cell is assumed to $L=30$. The solid-blue and dashed-red contours in (b) and (c) denote the envelopes of the resonance, respectively. Parameters: $\mu=5t_1$, $\xi=t_1$, $\kappa=0.006t_1$, and $t_2=0.7t_1$ for (b), and $t_2=1.3t_1$ for (c).}
	\label{fig:2}
\end{figure}

Based on the above topological invariants, there is a more transparent way to characterize the band topology, that is, by counting the number of times that $e^{{\rm i}\varphi(k)}$ wraps around the origin for $k\in[0,2\pi)$. Figure\,\ref{fig:1}(c) plots the circle of the function $q(k)$ centered at $(t_{1},0)$ in the complex plane. The winding number $\nu$ is non-vanishing and the phase $\varphi(k)$ is surjective to $[0,2\pi)$ once the origin is enclosed by $q(k)$. For the trivial winding number ($\nu=0$), the origin is not inside the circle of $q(k)$, which corresponds to the nonsurjection of $\varphi(k)$. Explicitly, the phase $|\varphi(k)|<\pi/2$ in $k\in[0,2\pi)$ for $\nu=0$.

We note that such a topological excitation can be measured in experiments. To do this, we define the correlation function $C_j(t)=-{\rm i}\Theta(t)\langle[\hat{x}_j(t),\hat{x}_j(0)]\rangle$ of a unit cell operator $\hat{x}_{j} = \sum_{s=\pm} (\hat{a}_{j,s} + \hat{a}_{j,s}^\dagger) /\sqrt{2}$. Here $\Theta$ is the Heaviside step function. In the periodic boundary condition, $C_j(t)$ does not depend on the index $j$. Transformed into the frequency domain, the correlation function is obtained by $C_j[\omega]=\frac{1}{2L}\sum_{k}(C_{k,+}[\omega]+C_{k,-}[\omega])$ ($L$ is the number of unit cells), where  
\begin{align}
C_{k,\pm}[\omega]=\frac{\cosh 2r\pm\cos\varphi(k)}{\omega-E_{\pm}(k)+{\rm i}\kappa},
\label{eq:Correlation}
\end{align}
and $\kappa^{-1}$ depicts the lifetime of the particle. Note that we have neglected the non-resonant terms. 

The imaginary part of the correlation function $-{\rm Im}C_j[\omega]$ reflects the resonance of the excitation spectrum of the system. More precisely, $-{\rm Im}C_{k,+}[\omega]$ and $-{\rm Im}C_{k,-}[\omega]$ exhibit the resonance of the upper band $E_{+}(k)$ and lower band $E_{-}(k)$ at momentum $k$, respectively, since $-{\rm Im}C_{k,\pm}[\omega] \propto \delta(\omega-E_{\pm}(k))$ in the limit $\kappa \rightarrow 0^{+}$. It implies that, for sufficiently small $\kappa$, each peak of $-{\rm Im}C_{j}[\omega]$ represents the resonance of frequency $\omega$ with the bands at momenta $k$, i.e., $\omega\simeq E_{\pm}(k)$ [see Figs.\,\ref{fig:2}(b) and \ref{fig:2}(c)]. These resonance peaks one-to-one correspond to $k\in[0,\pi]$ since $E_{\pm}(k)$ change monotonically and $C_{k,\pm}[\omega] = C_{-k,\pm}[\omega]$.

More importantly, the envelope of the resonance of $-{\rm Im}C_{j}[\omega]$ carries the topological information of the system through the numerators $\cosh 2r \pm \cos\varphi(k)$. Figure\,\ref{fig:2}(a) plots $\cos \varphi(k)$ for the trivial (solid-blue contour) and nontrivial (dashed-red contour) winding number. When $\nu=1$, the function $\cos \varphi(k)$ monotonically changes from $1$ to $-1$ for $k\in[0,\pi]$. We thus expect that the resonance envelope manifests the monotonicity for the high-lying topological excitation. Such a result can be verified by the dashed-red contour plotted in Fig.\,\ref{fig:2}(c). In contrast, $\cos\varphi(k)$ nearly equals to $1$ in $k\in[0,\pi]$ and $\varphi(0)=\varphi(\pi)=0$ when $\nu=0$. Therefore, as shown in Fig.\,\ref{fig:2}(b), the envelope (solid-blue contour) presents the non-monotonic behavior. As a summary, one can measure the monotonicity of the resonance envelope of a Bloch band to examine the topological or trivial bosonic Bogoliubov excitation for the 1D model. In addition, this monotonicity also reflects the Zak phase of a 1D system with inversion symmetry~\citep{Goren2018PRB}.

\subsection{Edge excitation and robustness against symmetry-preserving disorders}
\begin{figure}
	\begin{centering}
		\includegraphics[width=8.5cm]{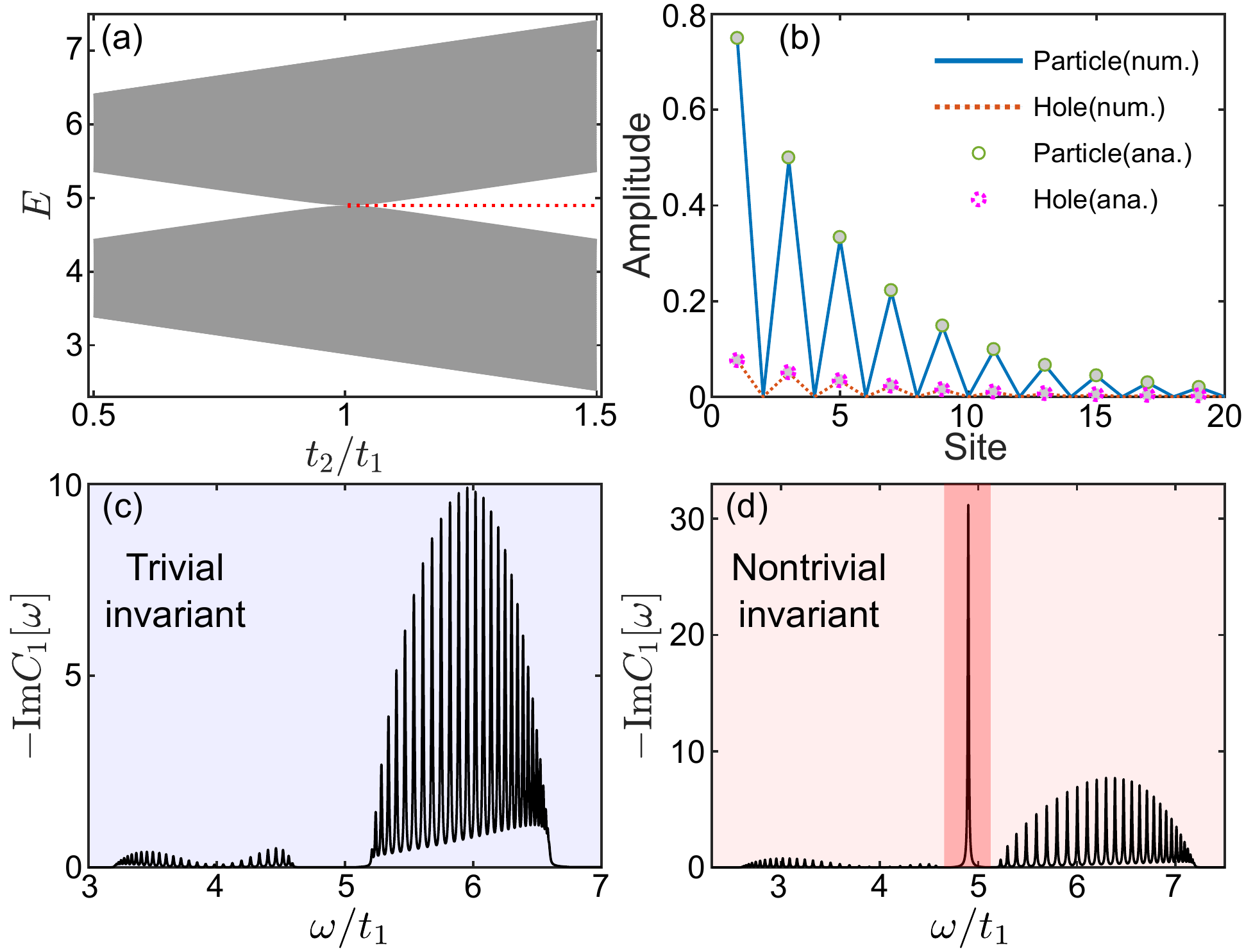}
	\end{centering}
	\caption{(a) Excitation spectrum $E$ versus $t_2$ in the open boundary condition with $L=100$ and $\mu=5t_1$. (b) Amplitude of the left edge-excitation operator, corresponding to the dashed-red line in (a). The circles denote the analytical solution given by Eq.\,(\ref{eq:SSH_EM}) while the lines are numerically plotted. The solid (dashed) lines and circles are responsible for the particle (hole) parts of the edge mode. (c),(d) $-{\rm Im}C_1[\omega]$ versus $\omega$ for the trivial and nontrivial topological invariants. The peak occurred at the midgap in (d) represents the resonance of the left edge-excitation operator. The related parameters are $L=30$, $\mu=5t_1$, $\xi=t_1$, $\kappa=0.006t_1$, and $t_2=0.7t_1$ for (c), and $t_2=1.3t_1$ for (d).}
	\label{fig:3}
\end{figure}

The bulk-edge correspondence ensures the occurrence of the edge excitation in the bulk gap when the open boundary condition is applied, as shown in Fig.\,\ref{fig:3}(a). Performing the condition to the system, we adopt the ansatz for the edge-excitation operators as
\begin{align}
\hat{\beta}_{{ L}}=\sum_{j=1}^{L}\delta^{j}\hat{\beta}_{j+},\ \ \hat{\beta}_{{ R}}=\sum_{j=1}^{L}\delta^{L-j+1}\hat{\beta}_{j-},
\label{eq:SSH_EM}
\end{align}
where $\delta=-t_1/t_2$ ($t_{1}<t_{2}$). Here the subscript $L/R$ denotes the left/right chirality of the system. It turns out that these edge-excitation operators $\hat{\beta}_{L}$ and $\hat{\beta}_{R}$ satisfy
\begin{align}
[\hat{\beta}_{{L}},\hat{H}_{{\rm I}}^{(\rm OBC)}] & =\tilde{\mu}\hat{\beta}_{{L}}+{{\cal O}}\left(\delta^{L+1}\right),\notag\\
[\hat{\beta}_{{R}},\hat{H}_{{\rm I}}^{(\rm OBC)}] & =\tilde{\mu}\hat{\beta}_{{R}}+{{\cal O}}\left(\delta^{L+1}\right),
\label{eq:SSH_Communtation}
\end{align}
where $\hat{H}_{\rm I}^{(\rm OBC)}$ denotes the Hamiltonian of the system in the open boundary condition. The last terms of Eq.\,(\ref{eq:SSH_Communtation}) are vanishing in the thermodynamic limit ($L\rightarrow\infty$). 

In the original basis, the edge-excitation operators are formed by the ``particles'' ($\hat{a}_{j,s}$) and ``holes'' ($\hat{a}^\dagger_{j,s}$). For example, $\hat{\beta}_{L}=\sum_{j}(\cosh r\delta^{j-1}\hat{a}_{j,+}+e^{\rm i\phi}\sinh r\hat{a}_{j,+}^\dagger)$. Figure \ref{fig:3}(b) plots the amplitude of the left edge-excitation operator for $L=100$. The numerical result is consistent with our analytical solution. Moreover, the edge excitations can also be measured by the correlation function $C_{j=1}[\omega]$ at the left-end unit cell. We note that this correlation-function approach can also be utilized to identify the signature of non-Hermitian (or spectral) topology within dissipative settings, such as the bosonic analog of Majorana zero mode~\citep{Flynn2021PRL}. As shown in Figs.\,\ref{fig:3}(c) and \ref{fig:3}(d), $-{\rm Im}C_{1}[\omega]$ has a peak at $\omega=\tilde{\mu}$ for $\nu=1$ while absent for $\nu=0$. Such a resonance peak indicates the occurrence of the left edge-excitation operator $\hat{\beta}_{L}$. Notably, here the resonance envelope is different from the case shown in Figs.\,\ref{fig:2}(b) and \ref{fig:2}(c) due to the effect of the open boundary condition.

\begin{figure}
	\begin{centering}
		\includegraphics[width=8.5cm]{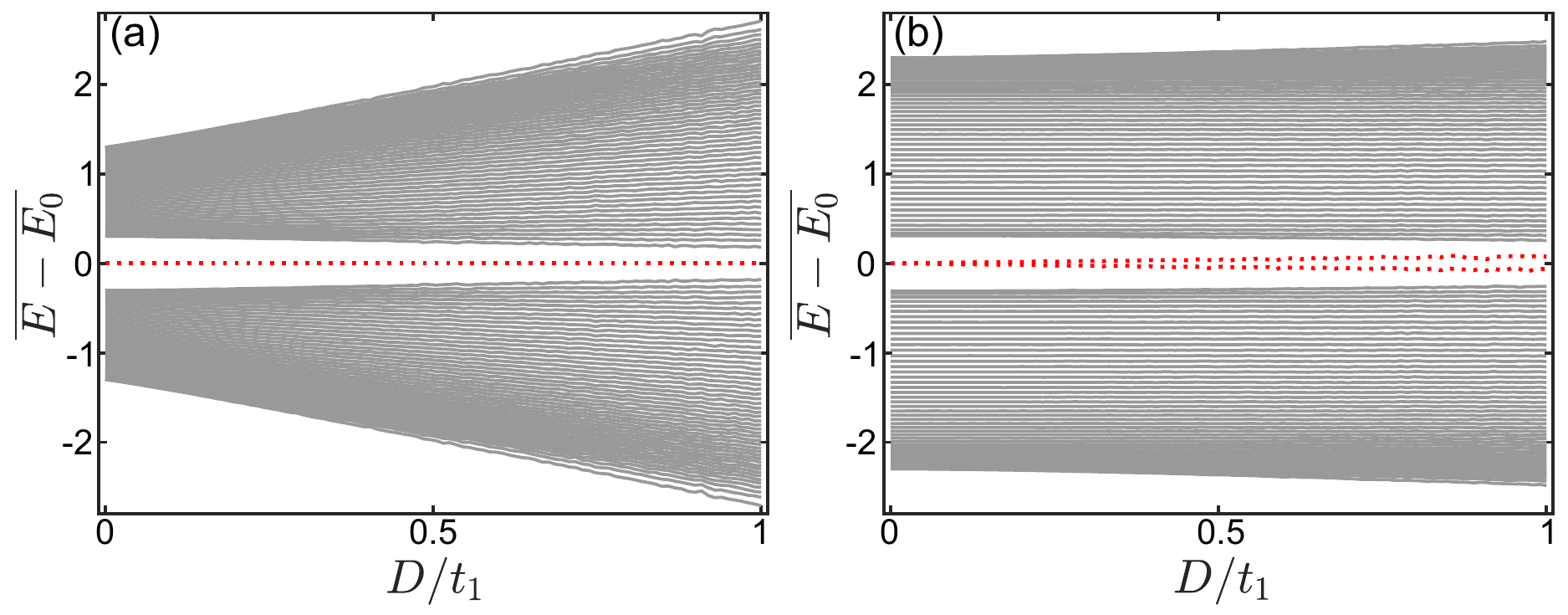}
	\end{centering}
	\caption{(a),(b) The mean excitation spectra $\overline{E}$ of 100 samples as a function of the disorder strength $D$ in the open boundary condition. The dashed-red lines in the midgap denote the edge modes in the presence of disorders. Here $E_0$ is the center of the excitation spectrum. In (a), the disorder is in the hopping terms of the samples while it is in onsite-energy terms in (b). The disorder is assumed to range $[-D,D]$. The parameters are $L=50$, $\mu=5t_1$, $\xi=t_1$, $t_2=1.3t_1$.}
	\label{fig:4}
\end{figure}

Due to the topological protection, the existence of such edge-excitation operators is robust against symmetry-preserving disorders. Here let us discuss the influence of two types of disorders on the edge excitations for comparison, i.e., hopping and onsite disorders. The former type preserves the sublattice symmetry and time-reversal symmetry. In this case, the edge-excitation operators still occur at the midgap and are modified to
\begin{align}
\hat{\beta}^{\rm (dis)}_{L}=&\sum_{j=1}^{L}\left(\prod_{l=1}^{j}-\frac{t_{2l-1}}{t_{2l}}\right)\hat{\beta}_{j,+},\notag\\
\hat{\beta}^{\rm (dis)}_{R}=&\sum_{j=1}^{L}\left(\prod_{l=L}^{L-j+1}-\frac{t_{2l-1}}{t_{2l}}\right)\hat{\beta}_{j,-},
\end{align}
where $t_{2l-1}$ and $t_{2l}$ are the strengths of intra- and intercell hoppings at the $l$-th unit cell. The two edge-excitation operators commute with the disordered Hamiltonian when $L$ is sufficiently large. It indicates that the edge excitations are robust against the symmetry-preserving disorder. Figure\,\ref{fig:4}(a) shows that the edge excitations remain unchanged even in the presence of strong disorder with strength $D$, which is consistent with our analysis. In contrast, the presence of onsite disorder breaks the sublattice symmetry. It, therefore, leads to the disappearance of edge excitations. In Fig.\,\ref{fig:4}(b), the edge excitations become non-degenerate and gradually join into the bulk without closing the band gap when the strength of disorder $D$ becomes strong. This contrast verifies that the high-lying topological bosonic Bogoliubov excitation (\ref{eq:Winding}) is intrinsically supported by the sublattice symmetry.

\section{Conclusions \label{sec:Conclusion}}

In conclusion, we have investigated the free-boson system governed by the bosonic BdG Hamiltonian  in the thermodynamic-stability regime (or slightly positive definite regime). The reduction of the particle-number-nonconserving BdG Hamiltonian to a single-particle one has been shown via a pseudo-unitary transformation. Building on this, we have unveiled the existence of the sublattice symmetry for the Bogoliubov excitation. Such a symmetry supplies an additional topological class, i.e., class AIII, which enriches the current framework for topological threefold way of the free-boson system. A category of bosonic BdG Hamiltonians respecting the sublattice symmetry has been presented. To further illustrate the topological Bogoliubov excitation of 1D free-boson system, we have also studied a prototypical model which has the nontrivial winding number (or symplectic polarization) guaranteed by this symmetry. A measurement of the topological invariant through the correlation function has been proposed. Moreover, as a topological effect, edge excitations have also been demonstrated, which are robust against symmetry-preserving disorders. Our work is expected to inspire the experimental activity since the proposed 1D model can be readily implemented in various platforms including the photonic system~\citep{Wang2021PRB}, superconducting circuit~\citep{Youssefi2022Nature}, optical lattice loaded with cold bosonic atoms~\citep{Wang2021Nature}, etc.

\section*{Acknowledgements}
We are grateful to Zixian Zhou for his insightful discussion on Lemma 1. This work is supported by the National Key Research and Development Program of China (Grant No.\,2021YFA1400700), and the National Natural Science Foundation of China (Grants No.\,11974125, No.\,11975103, and No.\,12205109).

\appendix
\section{Dynamical stability and its connection to thermodynamic stability \label{sec:DS}}

Here we discuss the dynamical stability for the free-boson system. A free-boson system is {\it dynamically stable} if the dynamical matrix $H_{\tau}({\bf k})$ is diagonalizable and its eigenvalues are real, i.e., there exists a pseudo-unitary matrix such that 
$$H_{\tau}({\bf k})V({\bf k})=V({\bf k})\Lambda({\bf k}),$$
where $\Lambda({\bf k})=E({\bf k})\oplus-E(-{\bf k})$ is a real and diagonal matrix. It implies that the normal mode of the dynamically stable $\hat{H}$ exhibits strictly bounded motion~\citep{Peano2018JMP, Flynn2020NJP}. Hence, the dynamical stability can be signaled by the complex eigenvalue or Jordan block.
For instance, the non-Hermiticity of $H_{\tau}({\bf k})$ allows the appearance of imaginary modes, which grow exponentially in time, indicating the dynamical instability. On the other hand, the ground state for the dynamically stable $\hat{H}$ is not a necessity because the system may have negative spectrum, which is called the Landau instability~\citep{Kawaguchi2012PR}. In the presence of Landau instability, the system can lower its energy by exciting a negative-eigenvalue mode.

At last, we remark that the two notions of thermodynamic and dynamical stability are independent of each other, although the system can be both thermodynamically and dynamically stable, for example, $H({\bf k})\geq 0$ and $H_{\tau}({\bf k})$ is diagonalizable. They can be distinguished by Jordan blocks and negative eigenvalues of the dynamical matrix in the diagonalization. In addition, the bosonic system {\it always} has a ground state in the thermodynamic-stability regime, which is in sharp contrast to the dynamical-stability regime.

\section{Details of time-reversal, particle-hole and chiral symmetries\label{sec:Sym_Def}}
In this appendix, we shall show more details of time-reversal, particle-hole and chiral symmetries on the many- and single-particle spaces for both of free-fermion and boson systems.

Let us define time-reversal, particle-hole and chiral symmetries implemented in free-fermion systems~\citep{Chiu2016RevModPhys}, and then these notions can be naturally inherited for free-boson systems. Firstly, let $\{\hat{\psi}_{{\bf k}i},\hat{\psi}_{{\bf k}i}^\dagger\}_{i=1}^{\tilde{N}}$ be a set of fermionic annihilation and creation operators with crystal momentum ${\bf k}$. Here $i$ denotes the non-spatial degree of freedom, e.g., Pauli-spin quantum number ($i=\pm 1/2$). These annihilation and creation operators satisfy the fermionic anti-commutation relation, $\{\hat{\psi}_{{\bf k}i},\hat{\psi}_{{\bf k}'j}^\dagger\}=\delta_{{\bf k}{\bf k}'}\delta_{ij}$. We consider a fermionic BdG Hamiltonian in momentum space,
\begin{equation}
\hat{H}_{f}=\sum_{{\bf k},ij}[\hat{\psi}_{{\bf k}i}^\dagger K_{ij}({\bf k})\hat{\psi}_{{\bf k}j}+\frac{1}{2}(\Delta_{ij}({\bf k})\hat{\psi}_{{\bf k}i}^\dagger\hat{\psi}_{-{\bf k}j}^\dagger+{\rm H.c.})],
\end{equation}
where $K({\bf k})=K^\dagger({\bf k})$ and $\Delta({\bf k})=-\Delta^T(-{\bf k})$ describes the pairing field in the free-fermion system. Writing into the Nambu representation, the Hamiltonian reads $\hat{H}_{f}=\frac{1}{2}\sum_{\bf k}[\hat{\Psi}^\dagger_{\bf k} H_{f}({\bf k})\hat{\Psi}_{\bf k}+{\rm Tr}K({\bf k})]$, where $\hat{\Psi}_{\bf k}=(\hat{\psi}_{{\bf k}1},\ldots,\hat{\psi}_{{\bf k}\tilde{N}},\hat{\psi}_{-{\bf k}1}^\dagger,\ldots,\hat{\psi}_{-{\bf k}\tilde{N}}^\dagger)^T$ is the fermionic Nambu spinor and 
$$H_{f}({\bf k})=\left(\begin{array}{cc}
K({\bf k})&\Delta({\bf k})\\ -\Delta^*(-{\bf k}) & -K^T(-{\bf k})
\end{array}\right)$$
is the BdG Hamiltonian on the single-particle (or particle-hole) space.

Time-reversal operator $\hat{T}$, which is antiunitary, acts on the fermionic operator, by definition, as
\begin{equation}
\hat{T}\hat{\psi}_{{\bf k}i}\hat{T}^{-1}=\sum_{j}(U_{T})_{ij}\hat{\psi}_{-{\bf k}j},\ \ \ \hat{T}{\rm i}\hat{T}^{-1}=-{\rm i},
\end{equation}
where $U_{T}$ is a unitary matrix. A system is time-reversal invariant if $\hat{T}$ preserves the fermionic anti-commutation relation, i.e., $\hat{T}\{\hat{\psi}_{{\bf k}i},\hat{\psi}_{{\bf k}j}^\dagger\}\hat{T}^{-1}=\delta_{ij}$, and if the Hamiltonian satisfies $\hat{T}\hat{H}_{f}\hat{T}^{-1}=\hat{H}_{f}$. It readily leads to the condition for the BdG Hamiltonian on the single-particle space
\begin{equation}
\hat{T}:\ \ {\cal T} H_{f}^{*}(-{\bf k}){\cal T}^{-1}=H_{f}({\bf k}),
\end{equation}
where the unitary matrix ${\cal T}=U_{T}^{\dagger}$ corresponds to the time-reversal operator on the single-particle space.

Particle-hole operator $\hat{C}$ is unitary and mixes the annihilation and creation operators,
\begin{equation}
\hat{C}\hat{\psi}_{{\bf k}i}\hat{C}^{-1}=\sum_{j}(U_{C}^{*})_{ij}\hat{\psi}_{-{\bf k}j}^\dagger,
\end{equation}
where $U_C$ is a unitary matrix. One can readily check that the anti-commutation relation is preserved under the particle-hole transformation, $\hat{C}\{\hat{\psi}_{{\bf k}i},\hat{\psi}_{{\bf k}j}^\dagger\}\hat{C}^{-1}=\delta_{ij}$. The system is invariant under particle-hole transformation if $\hat{C}\hat{H}_{f}\hat{C}^{-1}=\hat{H}_{f}$, which leads to
\begin{equation}
\hat{C}:\ \ {\cal C}H_{f}^T(-{\bf k}){\cal C}^{-1}=-H_{f}({\bf k}),
\end{equation}
where ${\cal C}=U_{C}^\dagger$ corresponds to the particle-hole operator on the single-particle space. Due to the Hermiticity of the BdG Hamiltonian $H_{f}^{*}=H_{f}^{T}$, we obtain the more familiar expression
\begin{equation}
{\cal C}H_{f}^*(-{\bf k}){\cal C}^{-1}=-H_{f}({\bf k}).
\end{equation}
Intrinsically, the BdG Hamiltonian considered here has the particle-hole symmetry with ${\cal C}=\tau_{1}$. 

Similarly, chiral operator, which is the combination of time-reversal and particle-hole operators, $\hat{\Gamma}=\hat{T}\hat{C}$, is antiunitary and also mixes the annihilation and creation operators,
\begin{equation}
\hat{\Gamma}\hat{\psi}_{{\bf k}i}\hat{\Gamma}^{-1}=\sum_{j}(U_{\Gamma}^*)_{ij}\hat{\psi}_{{\bf k}j}^\dagger,
\end{equation}
where $U_{\Gamma}=U_{C}^*U_{T}$. As a combination of time reversal and particle hole, this transformation also preserves the anti-commutator. And the chiral symmetry $\hat{\Gamma}\hat{H}_{f}\hat{\Gamma}^{-1}=\hat{H}_{f}$ yields the condition
\begin{equation}
\hat{\Gamma}:\ \ \Gamma H_{f}({\bf k})\Gamma^{-1}=-H_{f}({\bf k}),
\end{equation}
where $\Gamma=U_{\Gamma}^\dagger$. On the single-particle space, chiral operator $\hat{\Gamma}$ corresponds to the unitary matrix $\Gamma$.

Inherited from the notions of these symmetries for free fermions, one can define the bosonic version of time reversal $\hat{T}$, particle hole $\hat{C}$ and chiral $\hat{\Gamma}$. Note that the involved notations are also inherited for the free-boson case. These operators acting on the bosonic Nambu spinor $\hat{\Phi}_{{\bf k}}$ are defined by, respectively~\citep{Zhou2020JPA},
\begin{align}
\hat{T}\hat{\Phi}_{{\bf k}i}\hat{T}^{-1}&=\sum_{j}(U_{T})_{ij}\hat{\Phi}_{-{\bf k}j},\\
\hat{C}\hat{\Phi}_{{\bf k}i}\hat{C}^{-1}&=\sum_{j}(U_{C}^*)_{ij}\hat{\Phi}_{-{\bf k}j}^\dagger,\\
\hat{\Gamma}\hat{\Phi}_{{\bf k}i}\hat{\Gamma}^{-1}&=\sum_{j}(U_{\Gamma}^*)_{ij}\hat{\Phi}_{{\bf k}j}^\dagger,
\end{align}
where $U_T$, $U_{C}$ and $U_{\Gamma}=U_{C}^*U_{T}$ are unitary. Time-reversal and chiral operators are antiunitary ($\hat{T}{\rm i}\hat{T}^{-1}=-{\rm i}$ and $\hat{\Gamma}{\rm i}\hat{\Gamma}^{-1}=-{\rm i}$), while the particle-hole operator is unitary. To preserve the bosonic commutation relation (\ref{eq:CCR2}) under these transformations, i.e., $\hat{O}[\hat{\Phi}_{{\bf k}i},\hat{\Phi}_{{\bf k}j}^\dagger]\hat{O}^{-1}=(\tau_{3})_{ij}$, where $\hat{O}=\hat{T},\hat{C},\hat{\Gamma}$, the unitary matrices $U_T$, $U_{C}$ and $U_{\Gamma}$ should satisfy the following relations
\begin{equation}
[U_{T},\tau_{3}]=0,\ \ \{U_{C},\tau_{3}\}=0,\ \ \{U_{\Gamma},\tau_{3}\}=0.
\end{equation}
The three associated symmetries are then defined, $\hat{O}\hat{H}\hat{O}^{-1}=\hat{H}$, which leads to [ Eqs.\,(\ref{eq:TRS})-(\ref{eq:CS}) in the main text]
\begin{align}
&{\cal{T}}H_{\tau}^{*}(-{\bf k}){\cal{T}}^{-1}=H_{\tau}({\bf k}),\\
&{\cal{C}}H_{\tau}^{*}(-{\bf k}){\cal{C}}^{-1}=-H_{\tau}({\bf k}),\\
&\Gamma H_{\tau}({\bf k})\Gamma^{-1}=-H_{\tau}({\bf k}),
\end{align}
where ${\cal T}=U_{T}^\dagger$, ${\cal C}={U_{C}}^\dagger$ and $\Gamma=U_{\Gamma}^\dagger$. For the free-boson system considered here, the particle-hole operator exchanges the annihilation and creation operators, $\hat{C}:\hat{\Phi}_{{\bf k}}\rightarrow \tau_{1}(\hat{\Phi}_{-{\bf k}}^\dagger)^T=\hat{\Phi}_{\bf k}$, and the bosonic BdG Hamiltonian inherently has the particle-hole symmetry with the operator ${\cal C}=\tau_{1}$ on the Krein space.

\section{Proof of Lemma 1 \label{sec:Proof}}
Now, let us prove Lemma 1. Suppose the bosonic BdG Hamiltonian is dynamically stable. The dynamical matrix $H_{\tau}({\bf k})$ is thus diagonalizable and its eigenvalues are real, i.e., 
\begin{equation}
	H_{\tau}\left({\bf k}\right)V\left({\bf k}\right)=V\left({\bf k}\right)\Lambda\left({\bf k}\right),
\end{equation}
where $\Lambda\left({\bf k}\right)=E\left({\bf k}\right)\oplus-E\left(-{\bf k}\right)$ and $V\left({\bf k}\right)\tau_{3}V^{\dagger}\left({\bf k}\right)=\tau_{3}$, and ${\bf k}$ denotes the crystal momentum. We depict $E\left({\bf k}\right)$ ($-E\left(-{\bf k}\right)$) as the particle-(hole-)like spectrum of the free-boson system, due to the particle-hole symmetry. The columns of
\begin{equation*}
V({\bf k})=\left(\vert\phi_{1}({\bf k})\rangle,\ldots,\vert\phi_{\tilde{N}}({\bf k})\rangle,\vert\phi_{1}^{\prime}({\bf k})\rangle,\ldots,\vert\phi_{\tilde{N}}^{\prime}({\bf k})\rangle\right)
\end{equation*}
($\vert\phi_{i}^{\prime}({\bf k})\rangle=\tau_{1}\vert\phi_{i}^{*}(-{\bf k})\rangle$ for $1\leq i\leq\tilde{N}$)
are the eigenvectors of the dynamical matrix. They can span the indefinite Krein space $\mathbb{J}$: the former and latter $\tilde{N}$ eigenvectors have the positive and negative square norms, respectively,
\begin{align*}
\left\Vert \left|\phi_{i}\left({\bf k}\right)\right\rangle \right\Vert ^{2}=&\left\langle \phi_{i}\left({\bf k}\right)\left|\tau_{3}\right|\phi_{i}\left({\bf k}\right)\right\rangle>0,\\
\left\Vert \left|\phi_{i}^{\prime}\left({\bf k}\right)\right\rangle \right\Vert ^{2}=&\left\langle \phi_{i}^{*}\left(-{\bf k}\right)\left|\tau_{1}\tau_{3}\tau_{1}\right|\phi_{i}^{*}\left(-{\bf k}\right)\right\rangle<0.
\end{align*}
The Krein space can be decomposed to two subspaces, i.e., $\mathbb{J}=\mathbb{J}_{+}\oplus\mathbb{J}_{-}$, where $\mathbb{J}_{\pm}$ denote the subspaces of $\mathbb{J}$ equipped with positive and negative inner products, respectively. The former (latter) $\tilde{N}$ eigenvectors form a complete orthogonal normalized basis and can span a Hilbert space $\mathbb{H}$ with dimension $\tilde{N}$~\citep{Bognar1974}, corresponding to the particle (hole) space.

\subsection{Symmetry with unitary transformation}

Suppose the free-boson system have a symmetry, $OH_{\tau}\left({\bf k}\right)O^{-1}=\eta_{O}H_{\tau}\left({\bf k}\right)$, where the unitary matrix $O$ obeys $O\tau_{3}=\eta_{O}\tau_{3}O$ and $\eta_{O}=\pm$. Let $V_{j}\left({\bf k}\right)=\left(\left|\phi_{j_{1}}\left({\bf k}\right)\right\rangle, \left|\phi_{j_{2}}\left({\bf k}\right)\right\rangle, \ldots,\left|\phi_{j_{r}}\left({\bf k}\right)\right\rangle \right)$ be the $r$-fold eigenvectors of $H_{\tau}\left({\bf k}\right)$ with particle-like eigenvalue $E_{j}\left({\bf k}\right)\in\mathbb{R}$, i.e.,
\begin{equation}
H_{\tau}\left({\bf k}\right)V_{j}\left({\bf k}\right)=E_{j}\left({\bf k}\right)V_{j}\left({\bf k}\right).
\label{eq:A_eigenvec}
\end{equation}
It is worth noting that we have neglected the eigenvectors belonging to $\mathbb{J}_{-}$ with the same eigenvalue $E_j({\bf k})$ in Eq.\,(\ref{eq:A_eigenvec}), as the eigenvectors $V_j({\bf k})$ in $\mathbb{J}_{+}$ are fully capable of spanning the eigenspace $\mathbb{H}_{E_j}\subset\mathbb{H}$. Applying the unitary symmetry, we arrive at 
\begin{equation}
H_{\tau}\left({\bf k}\right)OV_{j}\left({\bf k}\right)=\eta_{O}E_{j}\left({\bf k}\right)OV_{j}\left({\bf k}\right).
\end{equation}

(i) For $\eta_{O}=+$, the columns of $OV_{j}\left({\bf k}\right)$ are still eigenvectors corresponding to the particle-like eigenvalue $E_{j}\left({\bf k}\right)$ and are complete for the eigenspace. It implies the linear combination
\begin{equation}
	OV_{j}\left({\bf k}\right)=V_{j}\left({\bf k}\right)L_{j}({\bf k}),
\end{equation}
where $L_{j}\left({\bf k}\right)$ is a $r$-dimensional matrix. Taking advantage of $V_{j}^{\dagger}\left({\bf k}\right)\tau_{3}V_{j}\left({\bf k}\right)=I_{r\times r}$, we obtain
\begin{align}
I_{r\times r} & =V_{j}^{\dagger}\left({\bf k}\right)O^{\dagger}\tau_{3}OV_{j}\left({\bf k}\right)\nonumber \\
& =L_{j}^{\dagger}V_{j}^{\dagger}\left({\bf k}\right)\tau_{3}V_{j}\left({\bf k}\right)L_{j}\left({\bf k}\right)\nonumber \\
& =L_{j}^{\dagger}\left({\bf k}\right)L_{j}\left({\bf k}\right).
\end{align}
That is, $L_{j}\left({\bf k}\right)$ is unitary. Similarly, we have
\begin{align}
OV'_{j}({\bf k})=V'_{j}({\bf k})L'_{j}({\bf k}),
\end{align}
where $V'_{j}({\bf k})=\tau_{1}V_{j}^{*}(-{\bf k})$ denote the eigenvectors with hole-like eigenvalue $-E_j(-{\bf k})$, and $L'_{j}({\bf k})$ is also unitary. Therefore, we have $OV\left({\bf k}\right)=V\left({\bf k}\right)L({\bf k})$, where $L\left({\bf k}\right)$ is unitary and takes the form
\begin{align}
L\left({\bf k}\right)=&L_{1}\left({\bf k}\right)\oplus L_{2}\left({\bf k}\right)\oplus\ldots\notag\\
&\ \ \oplus L'_{1}({\bf k})\oplus L'_{2}({\bf k})\oplus\ldots.
\end{align}
It gives us 
\begin{align}
OV\left({\bf k}\right)V^{\dagger}\left({\bf k}\right)O^{\dagger}=&V\left({\bf k}\right)L\left({\bf k}\right)L^{\dagger}\left({\bf k}\right)V^{\dagger}\left({\bf k}\right)\notag\\
=&V\left({\bf k}\right)V^{\dagger}\left({\bf k}\right).
\end{align}
Due to $e^{2W\left({\bf k}\right)}=V\left({\bf k}\right)V^{\dagger}\left({\bf k}\right)$, it finally arrives at
\begin{equation}
OW\left({\bf k}\right)O^{-1}=W\left({\bf k}\right).
\end{equation}

(ii) For $\eta_{O}=-$, $O$ exchanges the particle and hole mutually. $OV_{j}\left({\bf k}\right)$ corresponds to the hole-like eigenvalue $-E_{j}\left({\bf k}\right)=-E_{j^{\prime}}\left(-{\bf k}\right)$ with the $r^{\prime}$-fold eigenvectors $V_{j^{\prime}}^{\prime*}\left(-{\bf k}\right)$. We then have the linear combination
\begin{equation}
OV_{j}\left({\bf k}\right)=V_{j^{\prime}}^{\prime*}\left({\bf k}\right)R_{j^{\prime}}\left({\bf k}\right),
\end{equation}
\begin{align}
-I_{r\times r} & =V_{j}^{\dagger}\left({\bf k}\right)O^{\dagger}\tau_{3}OV_{j}\left({\bf k}\right)\nonumber \\
& =\left(R_{j^{\prime}}^{\dagger}\left({\bf k}\right)V_{j^{\prime}}^{\prime T}\left({\bf k}\right)\right)\tau_{3}\left(V_{j^{\prime}}^{\prime *}\left({\bf k}\right)R_{j^{\prime}}\left({\bf k}\right)\right)\nonumber \\
& =-R_{j^{\prime}}^{\dagger}\left({\bf k}\right)R_{j^{\prime}}\left({\bf k}\right),
\end{align}
where $R_{j^{\prime}}({\bf k})$ is a $r^{\prime}\times r$ rectangular matrix and $r^{\prime}\geq r$. Similarly, we can have
\begin{align}
OV_{j^{\prime}}^{\prime*}\left({\bf k}\right)=&V_{j}\left({\bf k}\right)R_{j}^{\prime}\left({\bf k}\right),
\end{align}
\begin{align}
I_{r^{\prime}\times r^{\prime}} =&V_{j^{\prime}}^{\prime T}\left({\bf k}\right) O^{\dagger}\tau_{3}O V_{j^{\prime}}^{\prime*}\left({\bf k}\right)\nonumber \\
=&R_{j}^{\prime\dagger}\left({\bf k}\right)V_{j}^{\dagger}\left({\bf k}\right)\tau_{3}V_{j}\left({\bf k}\right)R_{j}^{\prime}\left({\bf k}\right)\nonumber \\
=&R_{j}^{\prime\dagger}\left({\bf k}\right)R_{j}^{\prime}\left({\bf k}\right),
\end{align}
where $R_{j}^{\prime}\left({\bf k}\right)$ is a $r\times r^{\prime}$ matrix, and $r\geq r^{\prime}$. Thus, we obtain $r=r^{\prime}$, and $R_{j^{\prime}}\left({\bf k}\right)$ and $R_{j}^{\prime}\left({\bf k}\right)$ are unitary. Hence, we obtain $OV\left({\bf k}\right)=V\left({\bf k}\right)Z\left({\bf k}\right)$, where $Z({\bf k})$ is unitary and takes the form
\begin{align}
Z\left({\bf k}\right) & =\left(\begin{array}{cc}
		& R\left({\bf k}\right)\\
		R^{\prime}\left({\bf k}\right)
\end{array}\right),\\
R\left({\bf k}\right) & =R_{1^{\prime}}\left({\bf k}\right)\oplus R_{2^{\prime}}\left({\bf k}\right)\oplus\ldots,\notag\\	
R^{\prime}\left({\bf k}\right) & =R_{1}^{\prime}\left({\bf k}\right)\oplus R_{2}^{\prime}\left({\bf k}\right)\oplus\ldots.\notag
\end{align}
It, therefore, gives us
\begin{equation}
OW\left({\bf k}\right)O^{-1}=W\left({\bf k}\right).
\end{equation}
Hence, the mapping $e^{W({\bf k})}$ preserves unitary symmetries which commute or anti-commute with $\tau_{3}$.

\subsection{Symmetry with antiunitary transformation}
Now let us consider a symmetry with antiunitary transformation $OH_{\tau}^{*}\left(-{\bf k}\right)O^{-1}=\eta_{O}H_{\tau}\left({\bf k}\right)$, where the unitary matrix $O$ obeys $O\tau_{3}=\eta_{O}\tau_{3}O$. For the particle-like eigenvalue $E_{j}\left({\bf k}\right)$ with the $r$-fold eigenvectors $V_{j}\left({\bf k}\right)$, we have
\begin{equation}
H_{\tau}\left({\bf k}\right)OV_{j}^{*}\left(-{\bf k}\right)=\eta_{O}E_{j}\left(-{\bf k}\right)OV_{j}^{*}\left(-{\bf k}\right).
\end{equation}

(i) $\eta_{O}=+$. $OV_{j}^{*}\left(-{\bf k}\right)$ corresponds to particle-like eigenvalue $E_{l}\left({\bf k}\right)=E_{j}\left(-{\bf k}\right)$, with the eigenvectors $V_{l}\left({\bf k}\right)$. We have the linear combination
\begin{equation}
OV_{j}^{*}\left(-{\bf k}\right)=V_{l}\left({\bf k}\right){\cal L}_{l}\left({\bf k}\right).
\end{equation}
Based on the above analysis, it turns out ${\cal L}_{l}\left({\bf k}\right)$ is a $r$-dimensional unitary matrix. Similarly, we have
\begin{align}
OV_{j}^{\prime*}(-{\bf k})=V'_{l}({\bf k}){\cal L}'_{l}({\bf k}),
\end{align}
where ${\cal L}'_{l}({\bf k})$ is a $r$-dimensional unitary matrix. Thus, we have $OV^{*}\left(-{\bf k}\right)=V\left({\bf k}\right){\cal L}\left({\bf k}\right)$, where
\begin{align}
{\cal L}\left({\bf k}\right)=&{\cal L}_{1}\left({\bf k}\right)\oplus{\cal L}_{2}\left({\bf k}\right)\oplus\ldots\notag\\
&\ \ \oplus{\cal L}'_{1}\left({\bf k}\right)\oplus{\cal L}'_{2}\left({\bf k}\right)\oplus\ldots.
\end{align}
Thus, it also gives us 
\begin{equation}
OW^{*}\left(-{\bf k}\right)O^{-1}=W\left({\bf k}\right).
\end{equation}

(ii) For the case $\eta_{O}=-$, $OV_{j}^{*}\left(-{\bf k}\right)$ corresponds to the hole-like eigenvalue $-E_{j}\left(-{\bf k}\right)$, with the eigenvectors $V_{j}^{\prime}\left({\bf k}\right)$. Similarly, we still have the linear combinations
\begin{align}
OV_{j}^{*}\left(-{\bf k}\right) & =V_{j}^{\prime}\left({\bf k}\right){\cal R}_{j}\left({\bf k}\right),\\
OV_{j}^{\prime*}\left(-{\bf k}\right) & =V_{j}\left({\bf k}\right){\cal R}^{\prime}_{j}\left({\bf k}\right),
\end{align}
and ${\cal R}\left({\bf k}\right)$ and ${\cal R}^{\prime}\left({\bf k}\right)$ are $r$-dimensional unitary matrices. Equivalently, we have $OV^{*}\left(-{\bf k}\right)=V\left({\bf k}\right){\cal Z}\left({\bf k}\right)$,
where
\begin{align}
{\cal Z}\left({\bf k}\right) & =\left(\begin{array}{cc}
	& {\cal R}\left({\bf k}\right)\\
	{\cal R}^{\prime}\left({\bf k}\right)
\end{array}\right),\\
{\cal R}\left({\bf k}\right) & ={\cal R}_{1}\left({\bf k}\right)\oplus{\cal R}_{2}\left({\bf k}\right)\oplus\ldots,\notag\\
{\cal R}^{\prime}\left({\bf k}\right) & ={\cal R}_{1}^{\prime}\left({\bf k}\right)\oplus{\cal R}_{2}^{\prime}\left({\bf k}\right)\oplus\ldots\notag.
\end{align}
Thus, we have 
\begin{equation}
OW^{*}\left(-{\bf k}\right)O^{-1}=W\left({\bf k}\right).
\end{equation}

In a summary, we have proved that the mapping $e^{W\left({\bf k}\right)}$ preserves those non-unitary symmetries that commute or anti-commute with $\tau_{3}$, if the bosonic BdG Hamiltonian is dynamically stable.

\section{Topological triviality at zero energy \label{sec:No_go}}

Here we show that there is no nontrivial topology for the thermodynamically stable free-boson system. Concretely, we consider the case where the Bogoliubov bands of the system are gapped at zero energy. By performing the block-diagonalization (\ref{eq:Decomposition2}) with $e^{W({\bf k})}$, we obtain the transformed dynamical matrix $H'_{\tau}({\bf k})=\tilde{K}({\bf k})\oplus-\tilde{K}^{T}(-{\bf k})$ in the quasiparticle basis. Let $E_{j}({\bf k})>0$ and $\vert u_{j}({\bf k})\rangle$ with $j=1,\ldots,\tilde{N}$ be the eigenvalues and eigenstates of the single-particle Hamiltonian matrix $\tilde{K}({\bf k})$, respectively. Thus, based on Eq.\,(\ref{eq:Decomposition1}), we have 
\begin{align}
H'_{\tau}({\bf k})\vert \phi'_{j}({\bf k})\rangle =& E_{j}({\bf k})\vert \phi'_{j}({\bf k})\rangle, \notag\\ 
H'_{\tau}({\bf k})\tau_{1}\vert\phi_{j}^{\prime *}(-{\bf k})\rangle =&-E_{j}(-{\bf k})\tau_{1}\vert\phi_{j}^{\prime *}(-{\bf k})\rangle,
\label{eq:Eigen1}
\end{align}
where $\vert\phi'_{j}({\bf k})\rangle=(\vert u_{j}({\bf k})\rangle^{T},0)^{T}$.
Equation\,(\ref{eq:Eigen1}) can be expressed in a tight form, $\tilde{H}_{\tau}({\bf k})\tilde{U}({\bf k})=\tilde{U}({\bf k})\Lambda({\bf k})$,
with $\Lambda({\bf k})= E({\bf k})\oplus-E(-{\bf k})$
and 
\begin{equation}
V'=\left(\vert\phi_{1}\rangle,\ldots,\vert\phi'_{\tilde{N}}\rangle,\tau_{1}\vert\phi_{1}^{\prime*}\rangle,\ldots,\tau_{1}\vert\phi_{\tilde{N}}^{\prime *}\rangle\right).
	\label{eq:Unitary}
\end{equation}
Here we temporarily miss $({\bf k})$ for brevity. Note that the unitary matrix\,(\ref{eq:Unitary}) is formed of two blocks which correspond to the bands upper and lower than zero energy, respectively. Then the Bogoliubov bands can be continuously deformed to $\pm1$ without closing zero energy. Thus, the flattened Hamiltonian is given by 
\begin{align}
	H_{\tau{\rm flatten}}({\bf k}) =&V'({\bf k}) \left(\begin{array}{cc}
		I & 0\\
		0 & -I
	\end{array}\right)V^{\prime\dagger}({\bf k})\notag\\
=&\left(\begin{array}{cc}
		I & 0\\
		0 & -I
	\end{array}\right)=\tau_{3}.
	\label{eq:Flatten}
\end{align}
The second equality always holds due to the completeness of the set of $\{\vert u_{j}({\bf k})\rangle\}_{j=1}^{\tilde{N}}$. This flattened Hamiltonian implies that, in terms of the zero-energy gap, the system is \emph{always} topologically trivial in any symmetry classes and any dimensions~\citep{Lu2018arXiv, Xu2020PRB}.

For a free-boson system where the spectrum $\Lambda({\bf k})$ of the dynamical matrix $H_{\tau}({\bf k})$ is gapless at zero energy, we can obtain a corollary in light of the no-go theorem. That is, adding an infinitesimal perturbation $\Delta I$ ($\Delta\rightarrow0^{+}$) to $H({\bf k})$ does not influence the topology of the system since the approach does not result in the effect of the band inversion anyway even though a gap is open by the perturbation at zero energy. In other words, circumventing the possible exceptional point by the added energy $\Delta I$ does not affect the topology of the system at zero energy. Therefore, it is safe to study the topological excitation of the perturbed Hamiltonian $H({\bf k}) +\Delta I$ instead of $H({\bf k})$.

\section{Sublattice symmetry on the many-particle space \label{sec:SLS_MP}}
In general, it is sufficient to consider the sublattice symmetry at the single-particle level as the AZ classification is motivated to classify the properties of Bloch Hamiltonians in terms of non-unitary symmetries~\citep{Ryu2010NJP}. But, it would be better to help us understand the sublattice symmetry by lifting it on the many-particle space.

The sublattice operator acts on the bosonic Nambu spinor as
\begin{equation}
\hat{S}\hat{\Phi}_{{\bf k}i}\hat{S}^{-1}=\sum_{j}{\cal S}_{ij}\hat{\Phi}_{{\bf k}j},
\end{equation}
where ${\cal S}={\cal S}^\dagger={\cal S}^{-1}$. One can readily check that the bosonic commutation relation (\ref{eq:CCR2}) is preserved under the sublattice transformation, ${\hat{S}}[\hat{\Phi}_{{\bf k}i},\hat{\Phi}_{{\bf k}j}^\dagger]\hat{S}^{-1}=(\tau_{3})_{ij}$. In the squeezed-state representation with vacuum $\vert{\rm GS}\rangle$, the associated transformation becomes $\hat{S}:\tilde{\beta}_{{\bf k}}\rightarrow\tilde{{\cal S}}\tilde{\beta}_{{\bf k}}$. As the mapping $e^{W({\bf k})}$ preserves the sublattice symmetry proposed here, thus, the sublattice transformation also preserves the bosonic commutation relation, $\hat{S}[\tilde{\beta}_{{\bf k}i},\tilde{\beta}_{{\bf k}j}^\dagger]\hat{S}^{-1}=\delta_{ij}$.

We focus on the quadratic-bosonic Hamiltonian in the squeezed-state representation, which is given by $\hat{H}=\sum_{\bf k}\tilde{\beta}_{{\bf k}}^\dagger \tilde{K}({\bf k})\tilde{\beta}_{{\bf k}}$. As discussed in the main text, the sublattice symmetry requires the excitation spectrum being symmetric with respect to the nonzero energy $\epsilon$, and only the traceless part of the Hamiltonian matters for the band topology. We switch into the interaction picture with $\hat{H}_{0}=\epsilon\sum_{{\bf k}}\tilde{\beta}_{{\bf k}}^\dagger\tilde{\beta}_{{\bf k}}$, and obtain the desired Hamiltonian as $\hat{h}=\sum_{{\bf k}}\tilde{\beta}_{{\bf k}}^\dagger h({\bf k})\tilde{\beta}_{{\bf k}}$. Thus, the sublattice symmetry reads
\begin{equation}
\hat{S}\hat{h}\hat{S}^{-1}=-\hat{h}.
\end{equation}
It can be seen that this symmetry is not unitary on the many-particle space, which is different from the time-reversal, particle-hole and chiral symmetries.

\section{Proof on the quantization of the symplectic polarization \label{app:Derivation}}

Suppose that a 1D free-boson model with the Bloch BdG Hamiltonian matrix $H(k)$ respects the sublattice symmetry (${\cal S}$) with respect to the band gap $\epsilon$. Let $E_j(k)\geq0$ and $\vert \phi_{j}(k)\rangle=e^{W(k)}(\vert u_{j}(k)\rangle^T, 0)^T$ be the eigenvalues and right eigenstates of the dynamical matrix $H_{\tau}(k)$ below the band gap $\epsilon$, where $j=1,\ldots,\tilde{N}/2$. Due to the sublattice symmetry, we can obtain the eigenvalues $2\epsilon-E_j(k)$ above the gap. And the associated right eigenstates $\vert\tilde{\phi}_{j}(k)\rangle$ are given by
\begin{equation}
\vert\tilde{\phi}_{j}(k)\rangle=e^{W(k)}\left(\begin{array}{c}
\vert{\cal S}u_{j}(k)\rangle\\
0\\
\end{array}\right),
\end{equation}
which obey $H_{\tau}(k)\vert\tilde{\phi}_{j}(k)\rangle = (2\epsilon-E_j(k))\vert\tilde{\phi}_{j}(k)\rangle$. Subsequently, the symplectic polarization for the whole eigenstates can be calculated as 
\begin{widetext}
\begin{align}
	P^{{\rm whole}} &=\frac{\rm i}{2\pi}\sum_{j=1}^{N/2}\int_{\rm BZ}\left(\langle \phi_j(k)\vert\tau_3\vert{\rm d}\phi_j(k)\rangle+\langle\tilde{\phi}_j(k)\vert\tau_3\vert{\rm d}\tilde{\phi}_j(k)\rangle\right)\notag\\
	& =\frac{\rm i}{2\pi}\sum_{j=1}^{N/2}\int_{{\rm BZ}}\left[\left(\langle u_{j}(k)\vert,\ 0\right)e^{W(k)}\tau_{3}{\rm d}e^{W(k)}\left(\begin{array}{c}
		\vert u_{j}(k)\rangle\\
		0\end{array}\right)+\left(\langle{\cal S}u_{j}(k)\vert,\ 0\right)e^{W(k)}\tau_{3}{\rm d}e^{W(k)}\left(\begin{array}{c}
		\vert{\cal S}u_{j}(k)\rangle\\
		0\end{array}\right)\right]\notag\\
	&=\frac{\rm i}{2\pi}\sum_{j=1}^{N/2}\int_{{\rm BZ}}\left[\left(\langle u_{j}(k)\vert,\ 0\right) e^{-W(k)}{\rm d}e^{W(k)}\left(\begin{array}{c}
		\vert u_{j}(k)\rangle\\
		0\end{array}\right)+\left(\langle{\cal S}u_{j}(k)\vert,\ 0\right) e^{-W(k)}{\rm d}e^{W(k)}\left(\begin{array}{c}
		\vert {\cal S}u_{j}(k)\rangle\\
		0\end{array}\right)\right]\notag\\
	& =\frac{\rm i}{2\pi}\sum_{j=1}^{N/2}\int_{{\rm BZ}}\left( \langle u_{j}(k)\vert{\rm d}u_{j}(k)\rangle+\langle {\cal S } u_{j}(k)\vert{\rm d}{\cal S}u_{j}(k)\rangle\right)  =\frac{\rm i}{2\pi}\int_{{\rm BZ}}{\rm d}\log(\det U(k)),
\end{align}
\end{widetext}
where 
\[U(k)=(\vert u_1(k)\rangle,\ldots, \vert u_{\tilde{N}/2}(k)\rangle, {\cal S}\vert u_{1}(k)\rangle,\ldots, {\cal S}\vert u_{\tilde{N}/2}(k)\rangle)\] 
is unitary. 
In the fourth equality, we have used the identity 
\[
\left(\langle u_{j}(k)\vert,\ 0\right)\left(e^{-W(k)}\partial_{k}e^{W(k)}\right)\left(\begin{array}{c}
	\vert u_{j}(k)\rangle\\
	0\end{array}\right)=0.
\]
Therefore, $P^{{\rm whole}}$ is an integer $m\in\mathbb{Z}$
given by the winding number of the phase $-\arg(\det U(k))$ around the origin. Finally, the symplectic polarization of the states below energy $\epsilon$ is given by 
\begin{equation}
	P=\frac{\rm i}{2\pi}\sum_{j=1}^{N/2}\int_{{\rm BZ}}\langle u_{j}(k)\vert{\rm d}u_{j}(k)\rangle=\frac{P^{\rm whole}}{2}=\frac{m}{2}.
\end{equation}
Note that $\langle{\cal S}u_{j}(k)\vert{\rm d}{\cal S}u_{j}(k)\rangle=\langle u_{j}(k)\vert{\rm d}u_{j}(k)\rangle$. It can be seen that the symplectic polarization $P$ is quantized in units of $1/2$.


\begin{thebibliography}{99}
\bibitem [{Con()}]{Continuity}%
	\BibitemOpen
	\href@noop {} {}\bibinfo {note} {It is worth noting that for the
		positive-definite case, the continuity for the mapping is automatically
		satisfied if the dynamical matrix holds true.}\BibitemShut {Stop}%
\bibitem [{Exc()}]{Exception}%
	\BibitemOpen
	\href@noop {} {}\bibinfo {note} {Mathematically, the dynamical matrix
		$H_{\tau}$ is defective at this point and has a Jordan block after
		diagonalization.}\BibitemShut {Stop}%
\end{thebibliography}

%

\end{document}